\documentclass[12pt]{article}
\usepackage{a4wide,amssymb}
\pdfoutput=1
\usepackage{a4wide,amssymb,graphicx}
\usepackage{epsfig}
\usepackage[usenames,dvipsnames]{color}
\usepackage{slashed}
\newcommand{\be}{\begin{equation}}
\newcommand{\ee}{\end{equation}}
\newcommand{\bea}{\begin{eqnarray}}
\newcommand{\eea}{\end{eqnarray}}

\def\circa#1{\,\raise.3ex\hbox{$#1$\kern-.75em\lower1ex\hbox{$\sim$}}\,}

\begin{document}

\begin{titlepage}
%
%


%

\begin{centering}
\vspace{1cm}
{\Large {\bf  Inflection point inflation and reheating }} \\

\vspace{1.5cm}

{\bf Soo-Min Choi and Hyun Min Lee$^*$}
\vspace{.5cm}

{\it Department of Physics, Chung-Ang University, 06974 Seoul, Korea.} 
\\

\end{centering}
\vspace{2cm}

\begin{abstract}
\noindent
We revisit the inflection point inflation with an extended discussion to large field values and  consider the reheating effects on the inflationary predictions. Parametrizing the reheating dynamics in terms of the reheating temperature and the equation of state during inflation, we show how the observationally favored parameter space of inflection point inflation is affected by reheating dynamics. 
Consequently, we apply the general results to the inflation models with non-minimal coupling, such as the SM Higgs inflation and the $B-L$ Higgs inflation.

\end{abstract}

\vspace{5cm}

\begin{flushleft}
$^*$Email: hminlee@cau.ac.kr 
\end{flushleft}

\end{titlepage}

\section{Introduction}

Cosmic inflation is the main paradigm for the early Universe that solves horizon, homogeneity, isotropy and relic problems in Standard Big Bang Cosmology and it predicts the anisotropies of Cosmic Microwave Background(CMB) and seeds the structure formation in the early Universe. Inflation is based on the initial stage of an exponential expansion in the presence of a slow-rolling scalar field with nonzero vacuum energy. 
Inflaton has been assumed to be a SM singlet and have an almost flat potential during inflation. 
Since the temperature of  the Universe became almost zero after inflation ended, it is necessary to reheat the Universe by introducing the inflaton couplings to the SM particles.  As the inflaton is a SM singlet, the reheating mechanism \cite{reheating1} is mostly unknown and it could have been more complicated than we naively think. In most cases, taking a range of number of efoldings required for inflation between $N_*=50$ and $N_*=60$ is attributed to the uncertainties in reheating dynamics.
The recent results from Planck satellite constrain the spectral index with unprecedented precision \cite{planck} and the tensor-to-scalar ratio is best constrained by Planck $+$ BICEP2 $+$ Keck \cite{tensor}. Therefore, it becomes more important to pin down the uncertainties due to reheating dynamics in each class of inflation models \cite{reheating2}.

We consider the general predictions for CMB anisotropies and the effects of reheating dynamics on them in the context of inflection point inflation models \cite{inflection,inflection2}.  Unlike small or large inflation models, the inflection point inflation is based on specific values of the inflaton for which the inflaton potential has a zero curvature and a small slope while it is dominated by a constant vacuum energy during inflation.  We keep up to the cubic terms near the inflection point by assuming that the higher order terms vanish due to small field excursion during inflation or for a symmetry reason of the full inflaton potential.
It turns out that the effective potential for inflection point inflation is a good approximation in most cases,  
as far as the full potential shows such a specific point of the field space.  

We divide our discussion into two cases, one with small field excursion and one with large field excursion.   The effects of reheating dynamics is translated into the change of the number of efoldings required for solving the horizon problem. Depending on the reheating temperature and the equation of state during reheating,  we constrain the parameter space of the effective inflaton couplings from the spectral index and the tensor-to-scalar ratio. 
Then, we also apply the general results to concrete models for inflection point inflation such as SM Higgs inflation and $B-L$ Higgs inflation with non-minimal coupling where the renormalization group(RG) running effects become important for inflation. 
Both inflation models belong to the case with a  large field excursion of the canonical inflaton. 
 As a consequence, we identify the effective couplings of the inflaton potential with the fundamental parameters of the given model and show the parameter space where the tensor-to-scalar ratio is sizable. Depending on the models, the reheating effects become important for inflationary predictions, in particular, for the $B-L$ Higgs inflation with a relatively small inflaton mass and a small Higgs-portal coupling.  

The paper is organized as follows. 
We first present the general predictions of inflation observables in inflection point inflation and give a general discussion on the effects of reheating on the inflection point inflation. 
Then, we apply the general results to SM Higgs inflation and $B-L$ inflation with possible ranges of reheating temperature and equation of state.
Finally, conclusions are drawn.



\section{Inflection point inflation}

In this section, we summarize the general results for inflection point inflation.  
Some results on inflection point inflation have been obtained in the literature in a particular limit with small fields, for instance, in the context of warped D-brane inflation models \cite{inflection}.
We extend those results to the case with a  large field excursion for which the inflationary predictions are different from the case with a small field excursion. 
There might be a concern on the initial condition in inflection point inflation because the inflaton should not have a large kinetic energy near the inflection point for generic initial conditions given away from the inflection point \cite{initialcond}. However, there might be dissipative effects of the inflaton alleviating the initial condition problem, as they could provide friction terms to set the initial condition for inflection point inflation \cite{dissipative}.

\begin{figure}
  \begin{center}
   \includegraphics[height=0.30\textwidth]{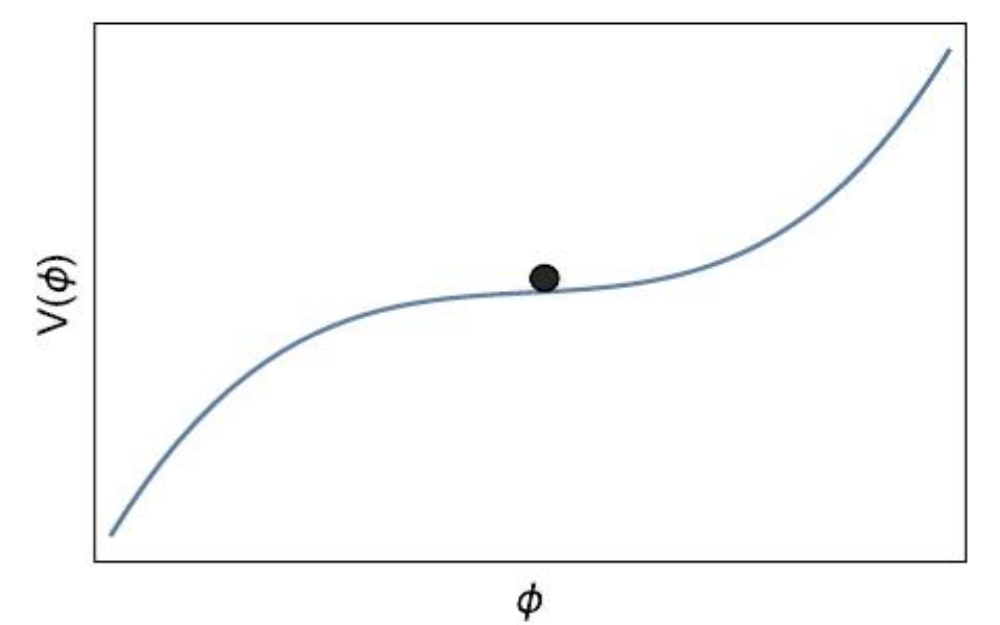}
    \end{center}
  \caption{Sketch of the inflaton potential near the inflection point.}
  \label{potential}
\end{figure}

We take the inflaton potential near the inflection point $\phi=\phi_0$ as
\be
V=V_0 +\lambda_1 (\phi-\phi_0) + \frac{1}{3!}\,\lambda_3 (\phi-\phi_0)^3
\ee
where $V_0,\lambda_1,\lambda_3$ are constant parameters depending on the models. 
A sketch of the inflaton potential in the above form is given in Fig.~\ref{potential}.
Henceforth, we work in units of the Planck mass set to $M_P=1$ and we take all the constant parameters above to be positive, without loss of generality \footnote{For $\lambda_1<0$, the inflection point is near a false vacuum so there is a problem with graceful exit. }.  
We assumed that higher order terms are suppressed due to a small field excursion during inflation or for some symmetry reason protecting the potential against them.

Then, the slow-roll parameters are computed as
\bea
\epsilon&=& \frac{1}{2}\left(\frac{\lambda_1+\frac{1}{2}\lambda_3 (\phi-\phi_0)^2}{V_0+\lambda_1 (\phi-\phi_0)+\frac{1}{6}\lambda_3 (\phi-\phi_0)^3} \right)^2, \label{epsilon}  \\
\eta &= & \frac{\lambda_3(\phi-\phi_0)}{V_0+\lambda_1 (\phi-\phi_0)+\frac{1}{6}\lambda_3 (\phi-\phi_0)^3}\, .  \label{eta}
\eea
 Moreover, from $N(\phi)=\int^\phi_{\phi_{\rm end}}\frac{d\phi}{\sqrt{2\epsilon}}$,  the number of efoldings from $\phi$ to the end of inflation is 
\bea
N(\phi)&=&\bigg(\frac{N_{\rm max}}{\pi}\,{\rm arctan}\Big(\sqrt{\frac{\lambda_3}{2\lambda_1}}\,(\phi-\phi_0) \Big)+\frac{1}{6}(\phi-\phi_0)^2 \nonumber \\
&&\quad +\frac{2\lambda_1}{3\lambda_3}\log(2\lambda_1+\lambda_3 (\phi-\phi_0)^2)\bigg)\bigg|^\phi_{\phi_{\rm end}}
\eea
with $N_{\rm max}\equiv \pi V_0\sqrt{\frac{2}{\lambda_1 \lambda_3}}$.
Choosing $\lambda_1/\lambda_3\lesssim 1$, we can use the approximate form for the number of efoldings as
\bea
N(\phi)\approx \bigg(\frac{N_{\rm max}}{\pi}\,{\rm arctan}\bigg(\sqrt{\frac{\lambda_3}{2\lambda_1}}\,(\phi-\phi_0)  \bigg)  +\frac{1}{6}(\phi-\phi_0)^2 \bigg)\bigg|^\phi_{\phi_{\rm end}}.
\eea
In this case, it is necessary to take $N_{\rm max}\gg 1$ for a large number of efoldings.
By taking $\phi=\phi_*$ at horizon exit with $|\phi_*-\phi_0|\lesssim 1$, the number of efoldings from horizon exit to the end of inflation is 
\be
N_*\approx \frac{N_{\rm max}}{\pi}\,\left({\rm arctan}\left(\frac{N_{\rm max}}{2\pi}\,\eta_* \right)-{\rm arctan}\left(\frac{N_{\rm max}}{2\pi}\,\eta_{\rm end} \right)\right)-\frac{1}{6}(\phi_{\rm end}-\phi_0)^2 \label{Nstar}
\ee
where $\eta_*\equiv \eta(\phi_*)$ and $\eta_{\rm end}\equiv \eta(\phi_{\rm end})$.
We note that for $\eta_*<2\pi/N_{\rm max}$ and $\eta_{\rm end}\sim -1$, the number of efoldings during inflation is $N_*\lesssim N_{\rm max}/2$. 
Consequently, from  $\eta_*\approx \frac{\lambda_3}{V_0}(\phi_*-\phi_0)$, we can solve eq.~(\ref{Nstar}) for $\phi_*$ as follows,
\bea
\phi_*-\phi_0 = \sqrt{\frac{2\lambda_1}{\lambda_3}} \tan\Theta  \label{phihe}
\eea
with
\bea
\Theta \approx\frac{\pi N_*}{N_{\rm max}} +\frac{\pi}{6N_{\rm max}}(\phi_{\rm end}-\phi_0)^2+\arctan\Big(\frac{ N_{\rm max}}{2\pi}\,\eta_{\rm end}\Big).  
\eea
Plugging eq.~(\ref{phihe}) into eqs.~(\ref{epsilon}) and (\ref{eta})  during inflation,  we obtain the slow-roll parameters at horizon exit as follows, 
\bea
\epsilon_* &=& \frac{1}{2}\Big(\frac{\lambda_1}{V_0}\Big)^2\left(\frac{\sec^2\Theta}{1+\frac{\lambda_1}{\lambda_3}\frac{2\pi}{N_{\rm max}}\tan\Theta+ \frac{\lambda_1}{\lambda_3}\frac{2\pi}{3N_{\rm max}}\tan^3\Theta }\right)^2, \\
\eta_* &=& \frac{2\pi}{N_{\rm max}}\frac{\tan\Theta}{1+\frac{\lambda_1}{\lambda_3}\frac{2\pi}{N_{\rm max}}\tan\Theta+ \frac{\lambda_1}{\lambda_3}\frac{2\pi}{3N_{\rm max}}\tan^3\Theta}.
\eea

On the other hand, inflation ends when $\epsilon_{\rm end}=1$. Thus, from $\epsilon_{\rm end}=1$ with eq.~(\ref{epsilon}), we can determine the inflaton field value at the end of inflation as
\bea
\phi_{\rm end}-\phi_0=\left\{\begin{array}{c}\frac{1}{\sqrt{2}}+(R+\sqrt{Q^3+R^2})^{1/3}+ (R-\sqrt{Q^3+R^2})^{1/3},\quad Q^3+R^2>0,\\
\frac{1}{\sqrt{2}}+2\sqrt{-Q}\cos\Big(\frac{1}{3}\arccos(R/\sqrt{-Q^3})+\frac{2\pi}{3}\Big), \quad Q^3+R^2\leq 0
\end{array}\right. \label{endsol}
\eea
where 
\bea
R&\equiv&  -\frac{3V_0}{\lambda_3} +\frac{1}{2\sqrt{2}}, \\
Q&\equiv & \frac{2\lambda_1}{\lambda_3}-\frac{1}{2}.
\eea
For $\lambda_1/\lambda_3\lesssim 1$, the inflaton field excursion in inflection point inflation is determined mainly by the value of $\lambda_3/V_0$. From the inflation field value at the end of inflation in eq.(\ref{endsol}), the inflaton makes a trans-Planckian excursion during inflation for $\lambda_3/V_0\lesssim 1$ while becoming sub-Planckian for $\lambda_3/V_0\gtrsim 1$.

Therefore, from $n_s=1-6\epsilon_*+2\eta_*$ and $r=16\epsilon_*$, the spectral index and the tensor-to-scalar ratio becomes
\bea
n_s&=&1-3\Big(\frac{\lambda_1}{V_0}\Big)^2\left(\frac{\sec^2\Theta}{1+\frac{\lambda_1}{\lambda_3}\frac{2\pi}{N_{\rm max}}\tan\Theta+ \frac{\lambda_1}{\lambda_3}\frac{2\pi}{3N_{\rm max}}\tan^3\Theta }\right)^2 \nonumber \\
&&+\frac{4\pi}{N_{\rm max}}\frac{\tan\Theta}{1+\frac{\lambda_1}{\lambda_3}\frac{2\pi}{N_{\rm max}}\tan\Theta+ \frac{\lambda_1}{\lambda_3}\frac{2\pi}{3N_{\rm max}}\tan^3\Theta}\,, \label{sindex} \\
r& =  & 8\Big(\frac{\lambda_1}{V_0}\Big)^2\left(\frac{\sec^2\Theta}{1+\frac{\lambda_1}{\lambda_3}\frac{2\pi}{N_{\rm max}}\tan\Theta+ \frac{\lambda_1}{\lambda_3}\frac{2\pi}{3N_{\rm max}}\tan^3\Theta }\right)^2. 
\label{tensor}
\eea
By using the above results, we find that the spectral index is correlated to the tensor-to-scalar ratio in this model as follows,
\be
n_s= 1-\frac{3}{8} \,r +\frac{1}{2}\sqrt{\frac{\lambda_3 r}{\lambda_1}}\, \sin(2\Theta).
\ee
Thus, for a fixed value of $\tan\Theta$ in eq.~(\ref{phihe}), we can correlate between the spectral index and the tensor-to-scalar ratio. 
We note that the measured spectral index and the bound on the tensor-to-scalar ratio, are $n_s=0.9652\pm 0.0047$ and $r<0.10$ at 95 $\%$ C.L., respectively, from Planck TT, TE, EE $+$ low P \cite{planck}.

Moreover, from the CMB normalization, $A_s=\frac{1}{24\pi^2}\,\frac{V_*}{\epsilon_*}\simeq 2.196\times 10^{-9}$, at the Planck pivot scale of $k=0.05\,{\rm Mpc}^{-1}$, we obtain
\bea
\frac{V^3_0}{\lambda^2_1}\cos^4\Theta\approx 2.60\times 10^{-7} .  \label{norm}
\eea
Then, the sufficient number of efoldings and the spectral index constrain the following combination of the model parameters,
\bea
\frac{V_0}{\lambda^2_3}\approx 2.05\times 10^{12}\left(\frac{N_{\rm max}}{120} \right)^4\cos^4\Theta.  \label{efolds}
\eea
Therefore, we can express the model parameters in terms of the tensor-to-scalar ratio $r$, $N_{\rm max}$ and $\cos\Theta$ as
\bea
V_0&\approx &3.25\times 10^{-9}\left(\frac{r}{0.10}\right), \label{v0}\\
\lambda_3&\approx& 3.98\times 10^{-11}\left(\frac{120}{N_{\rm max}}\right)^2\left(\frac{r}{0.10}\right)^{1/2}\sec^2\Theta, \label{lam3} \\
\lambda_1&\approx& 3.63\times 10^{-10} \left(\frac{r}{0.10}\right)^{3/2} \cos^2\Theta.  \label{lam10}
\eea

Furthermore, using eqs.~(\ref{v0}) and (\ref{lam3}), the useful discriminator between small and large field inflations, $\lambda_3/V_0$, is given by
\be
\frac{\lambda_3}{V_0}\approx\left(\frac{120}{N_{\rm max}} \right)^2 \left(\frac{1.50\times 10^{-5}}{r} \right)^{1/2} \sec^2\Theta. \label{discrim}
\ee
Therefore, for sub-Planckian inflation field excursions with $\lambda_3/V_0\gtrsim 1$, the tensor-to-scalar ratio becomes about $10^{-5}$ or less. On the other hand, for trans-Planckian field excursions, the tensor-to-scalar ratio as large as $r=0.1$ can be achieved for $\lambda_3/V_0\sim 10^{-2}$ or $|\phi_{\rm end}-\phi_*|\sim 10$.
From eqs.(\ref{lam3}) and (\ref{lam10}), we also find the ratio of $\lambda_1$ and $\lambda_3$ couplings as
\be
\frac{\lambda_1}{\lambda_3}\approx 1.37\times 10^{-3}\,\left(\frac{N_{\rm max}}{120} \right)^2\left(\frac{r}{1.50\times 10^{-5}}\right) \cos^4\Theta.
\ee
Then, for $N_{\rm max}\sim 100$ and $r \cos^4\Theta \lesssim 0.01$, we get $\lambda_1\lesssim \lambda_3$.

\begin{figure}
  \begin{center}
   \includegraphics[height=0.48\textwidth]{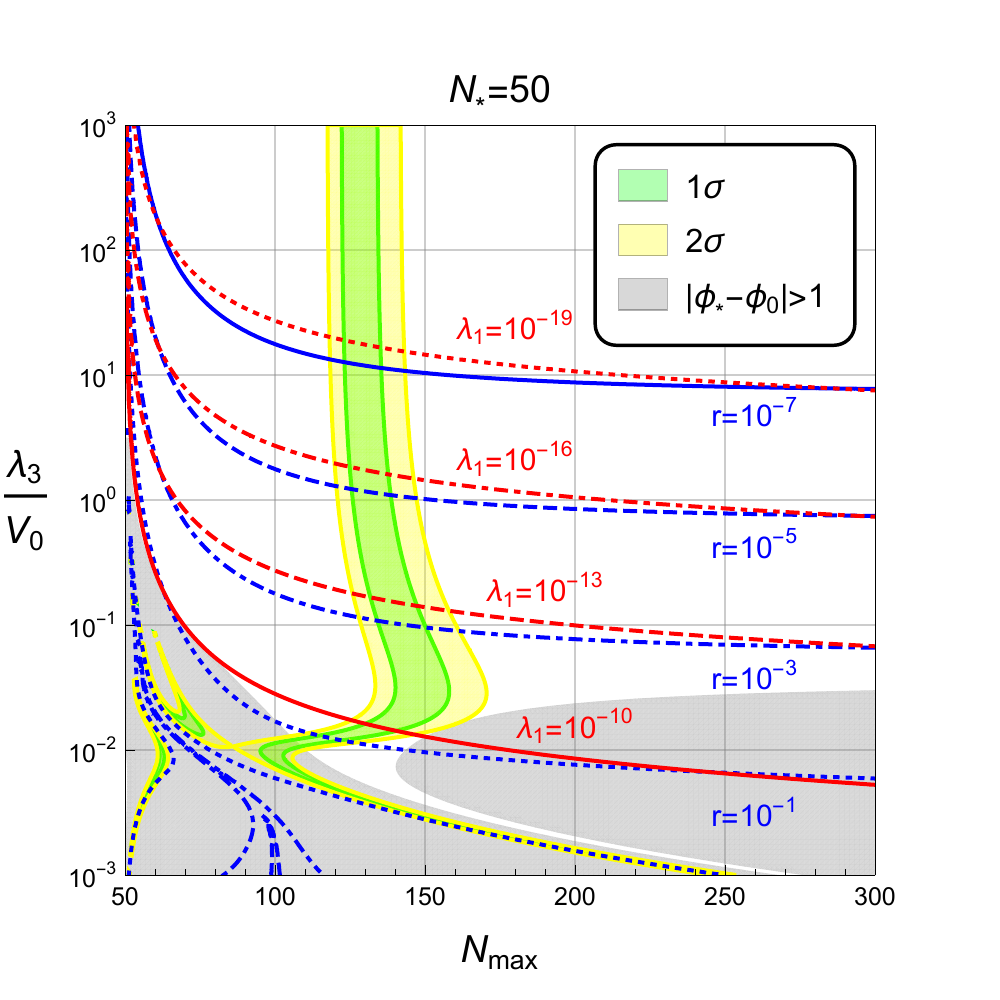}
   \includegraphics[height=0.48\textwidth]{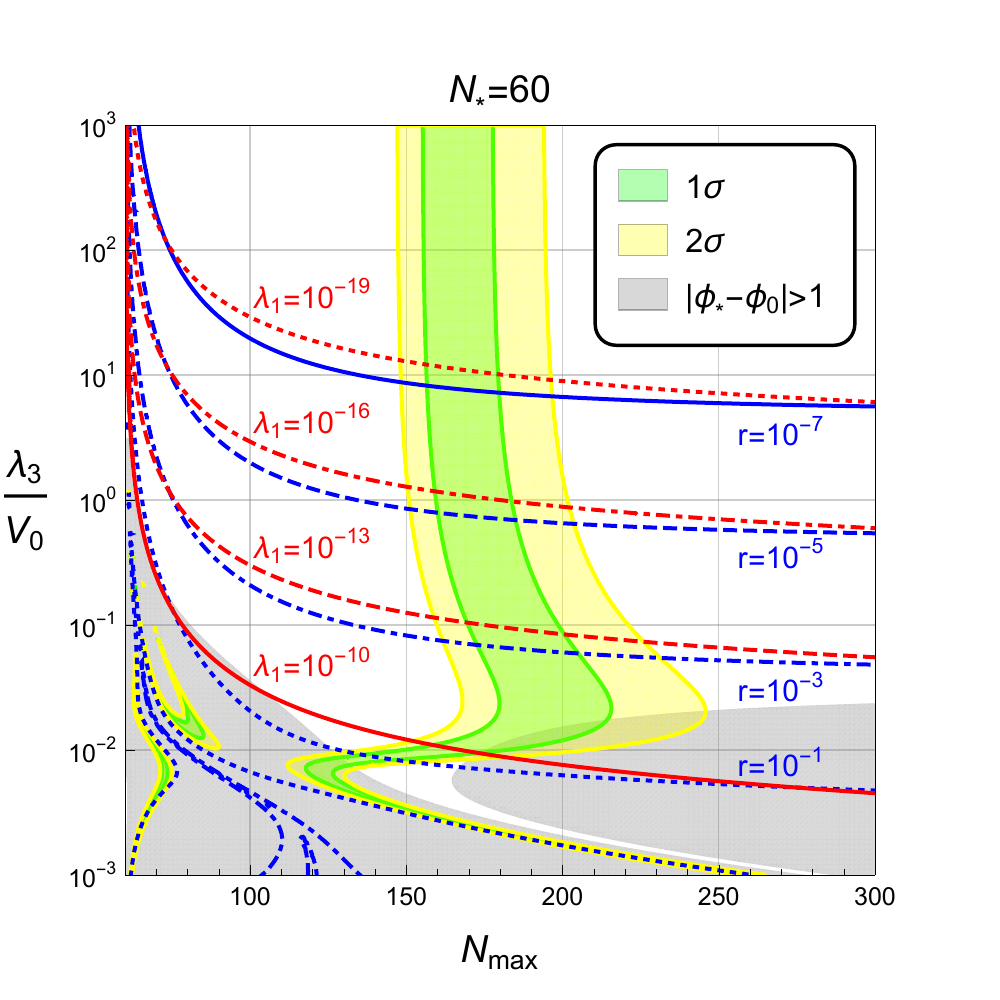}
   \end{center}
  \caption{Parameter space for $\lambda_3/V_0$ vs $N_{\rm max}\equiv \pi V_0\sqrt{2/(\lambda_1\lambda_3)}$, satisfying the spectral index measured by Planck within $1\sigma$ (green) and $2\sigma$ (yellow) for the number of efoldings, $N=50$ on left and $N=60$ on right. For a given value of $\lambda_1$, the CMB normalization is satisfied only along the red lines. Some values of tensor-to-scalar ratio, $r=0.1, 10^{-3}, 10^{-5}, 10^{-7}$, are shown in dotted, dot-dashed, dashed and solid lines (from bottom to top), respectively.  
  The region with inflation field value at horizon exit given by $|\phi_*-\phi_0|>1$ is shown in gray.}
  \label{noreheat}
\end{figure}

In Fig.~\ref{noreheat}, the parameter space for $\lambda_3/V_0$ vs $N_{\rm max}$ is shown, under the condition that the spectral index is obtained within $1\sigma$ and $2\sigma$ of the Planck data, depending on the number of efoldings, $N=50$ or $60$. For given ratios of parameters, $\lambda_3/V_0$ and $\lambda_1/V_0$, the CMB normalization (\ref{norm}) is achieved along a line in the parameter space in Fig. \ref{noreheat}. Representative values of $\lambda_1$, such as $\lambda_1=10^{-10}, 10^{-13}, 10^{-16}, 10^{-19}$, are overlaid in solid, dashed, dot-dashed and dotted red lines in the same figure. 
The boundary dividing the sub- and trans-Planckian inflaton excursions appears around $r\simeq 10^{-5}$ or $10^{-6}$. There is an isolated region for small $\lambda_3/V_0$ and a relatively small $N_{\rm max}$ in the left lower corner. In this region, however, our assumption that $|\phi_*-\phi_0|<1$, namely, the fact that the inflation starts taking place near the inflection point is violated, so we don't consider the region any more.

\subsection{Small-field case}

When $\lambda_3/V_0\gtrsim 1$, we have discussed that the field excursion during inflation is sub-Planckian. In this case, we can trust the perturbative expansion of the inflaton potential near the inflection point with higher order terms of $(\phi-\phi_0)$. In this case, the model parameters have to satisfy $\sqrt{\lambda_1\lambda_3}\ll V_0 \lesssim \lambda_3$, which corresponds to  the hierarchy, $\lambda_1\ll V_0 \lesssim \lambda_3$. 
Moreover, from eq.~(\ref{discrim}) with $N_{\rm max}\sim 100$ or as shown in the region with $|\phi_*-\phi_0|<1$ in Fig.~\ref{noreheat}, we find that $r\lesssim 10^{-5}$ or $\epsilon\lesssim 10^{-6}$ in the small-field case.
In this case, since $\epsilon_*\ll|\eta_*|$, we can ignore the contribution from $\epsilon_*$ in the spectral index. 
Then, from eq.~(\ref{sindex}) with $\eta_{\rm end}=-1$,  the spectral index becomes \cite{inflection}
\bea
n_s&=&1+\frac{4\pi}{N_{\rm max}}\,\tan\left(\pi\frac{N_*}{N_{\rm max}}-{\rm arctan}\left(\frac{N_{\rm max}}{2\pi} \right) \right).
\eea
Therefore, for $N_*=50(60)$, we have derived the bounds on $N_{\rm max}$ as $128<N_{\rm max}<141(162<N_{\rm max}<187)$ within $1\sigma$  from the Planck central value. This result can be seen clearly in Fig.~\ref{noreheat} for the region with $\lambda_3/V_0\gg 1$.

\subsection{Large-field case}

When $\lambda_3/V_0 \lesssim 1$, the inflaton can make a field excursion towards trans-Planckian values during inflation. Then, there is a concern about the higher order terms in the perturbative expansion near the inflection point. However, if the full inflaton potential is known to have higher order terms suppressed near the inflection point, the case with trans-Planckian field values remains an interesting possibility. 
 In this case, the spectral index is given by eq.~(\ref{sindex}).
Moreover, from eq.~(\ref{discrim}), a sizable tensor-to-scalar ratio can be obtained. 
For instance, for $r\sim 0.10$ and $N_{\rm max}\sim 100$, the model parameters have to satisfy a different hierarchy, $\lambda_3\ll \lambda_1\ll V_0$. 
As shown in Fig.~\ref{noreheat}, for $\lambda_3/V_0 \lesssim 1$ for which the inflaton field makes a trans-Planckian field excursion from the inflection point during inflation, the tensor-to-scalar ratio should be greater than $r\sim 10^{-5}$ and the linear term should be larger than $\lambda_1\sim 10^{-15}$ in the region where the observed spectral index is obtained.

\section{Reheating and number of efoldings}

We parametrize the reheating dynamics in terms of the reheating temperature, at which the inflaton energy is completely transferred into the radiation of the SM particles, and the equation of state during reheating, $w=p/\rho$ \cite{EOS}, which is restricted to $-\frac{1}{3}\leq w \leq 1$ for no more acceleration, ${\ddot a}\leq 0$. 
When the form of the inflaton potential is known after inflation, the equation of state might be in principle determined from the averaged energy density and pressure of the inflaton during reheating. 
However, it is known that a small inflaton coupling to spectator fields changes the equation of state by a large amount in a sufficiently short time scale during the process of preheating \cite{EOS}. Moreover, in some cases, the precise form of the inflaton potential near the oscillation does not affect the inflection point inflation, e.g., in the case of inflation models with non-minimal coupling as will be discussed in a later section. 
Therefore, we treat the equation of state to be a free parameter.

\begin{figure}
  \begin{center}
   \includegraphics[height=0.40\textwidth]{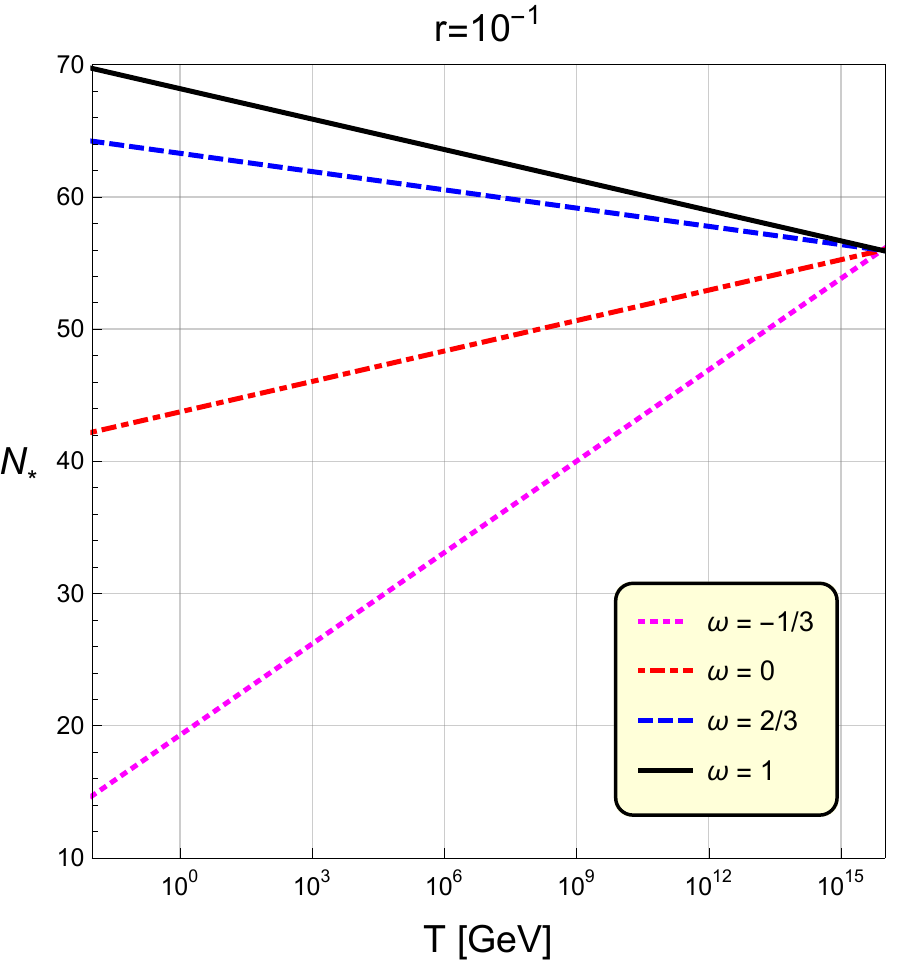}
   \includegraphics[height=0.40\textwidth]{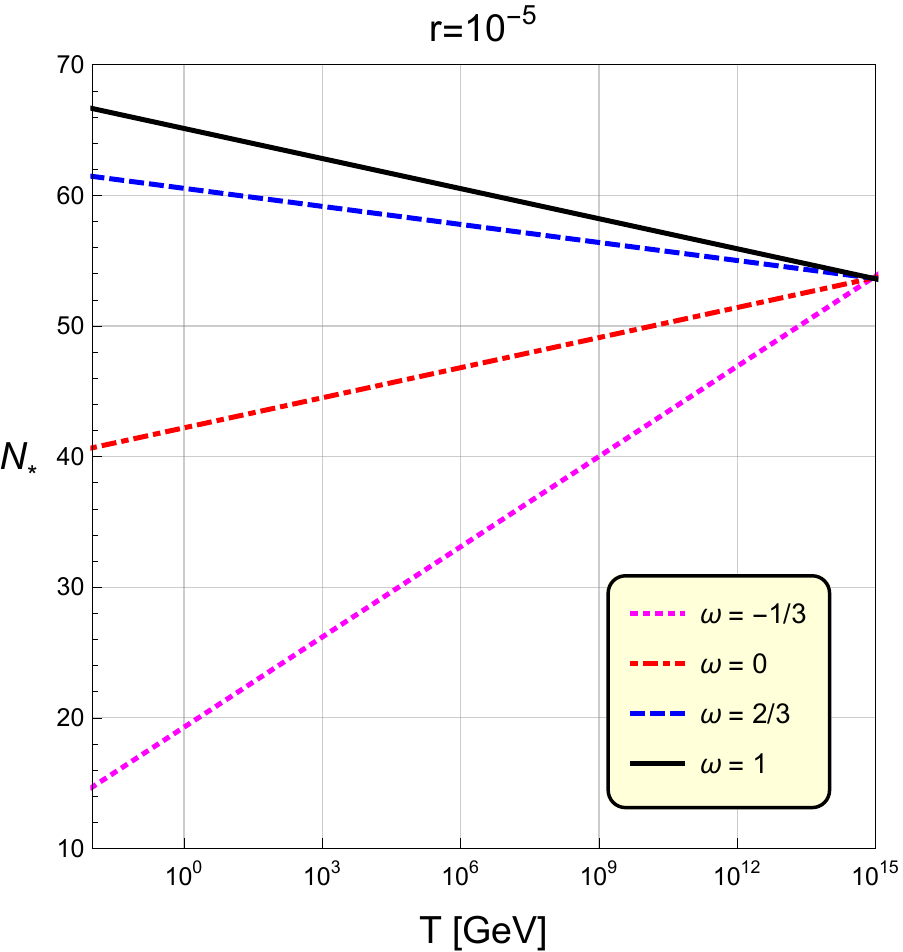} 
   \end{center}
  \caption{Number of efoldings $N_*$ during inflation as a function of the reheating temperature $T_{\rm rh}$. The equation of state during reheating is chosen to $w=-\frac{1}{3}, 0, \frac{2}{3}, 1$ are denoted by solid(red), dashed (green), dotted (green) and dot-dashed (magenta) lines (from bottom to top), respectively. The tensor-to-scalar ratio is chosen to $r=10^{-5}, 0.1$, on left and right figures, respectively. }
  \label{ncmb}
\end{figure}

Assuming that there is no entropy change between the reheating and the present,
we obtain the number of efoldings between the horizon exit and the end of inflation required to solve the horizon problem up to the scale of $k$, as follows,
\bea
N_* &=& 67.2 +\frac{1}{4}(3w-1) N_{\rm rh} -\ln\left(\frac{k}{a_0 H_0} \right) -\ln\left(\frac{V^{1/4}_{\rm end}}{H_*} \right) -\frac{1}{12} \ln(g_*(T_{\rm rh}))
\eea
where $g_*(T_{\rm rh})$ is the number of relativisitic species in thermal bath at reheating temperature $T_{\rm rh}$, and $N_{\rm rh}$ is the number of efoldings during reheating, given by
\be
N_{\rm rh} = \frac{1}{3(1+w)} \ln\left(\frac{45}{\pi^2}\frac{V_{\rm end}}{g_*(T_{\rm rh})T^4_{\rm rh}} \right).
\ee
Taking $V_{\rm end}\simeq V_*$ and choosing $g_*(T_{\rm rh})\simeq 100$ for $T_{\rm rh}\gtrsim 100\,{\rm GeV}$,  the formula for the number of efoldings becomes
\bea
N_*=61.4 +\frac{3w-1}{12(1+w)} \ln\left(\frac{45}{\pi^2}\frac{V_*}{g_*(T_{\rm rh})T^4_{\rm rh}} \right)-\ln\left(\frac{V^{1/4}_*}{H_*} \right). 
\eea
We note that the last term depends on the inflaton potential and the second term depends on both inflation models and reheating dynamics. In Fig.~\ref{ncmb}, the dependence of the number of efoldings on the reheating temperature and the equation of state during inflation is given.

In the case of an instantaneous reheating for which the reheating temperature is given by $T_{\rm rh}=(45 V_{\rm end}/\pi^2 g_*)^{1/4}\simeq (8.5\times 10^{14}\,{\rm GeV})(r/10^{-5})^{1/4}$, we would get $N_{\rm rh}=0$, so the number of efoldings depends only on the inflaton potential. In this case, the number of efoldings for the inflection point inflation is given by $N_*=53.7+\frac{1}{4}\ln(r/10^{-5})$.

\begin{figure}
  \begin{center}
   \includegraphics[height=0.30\textwidth]{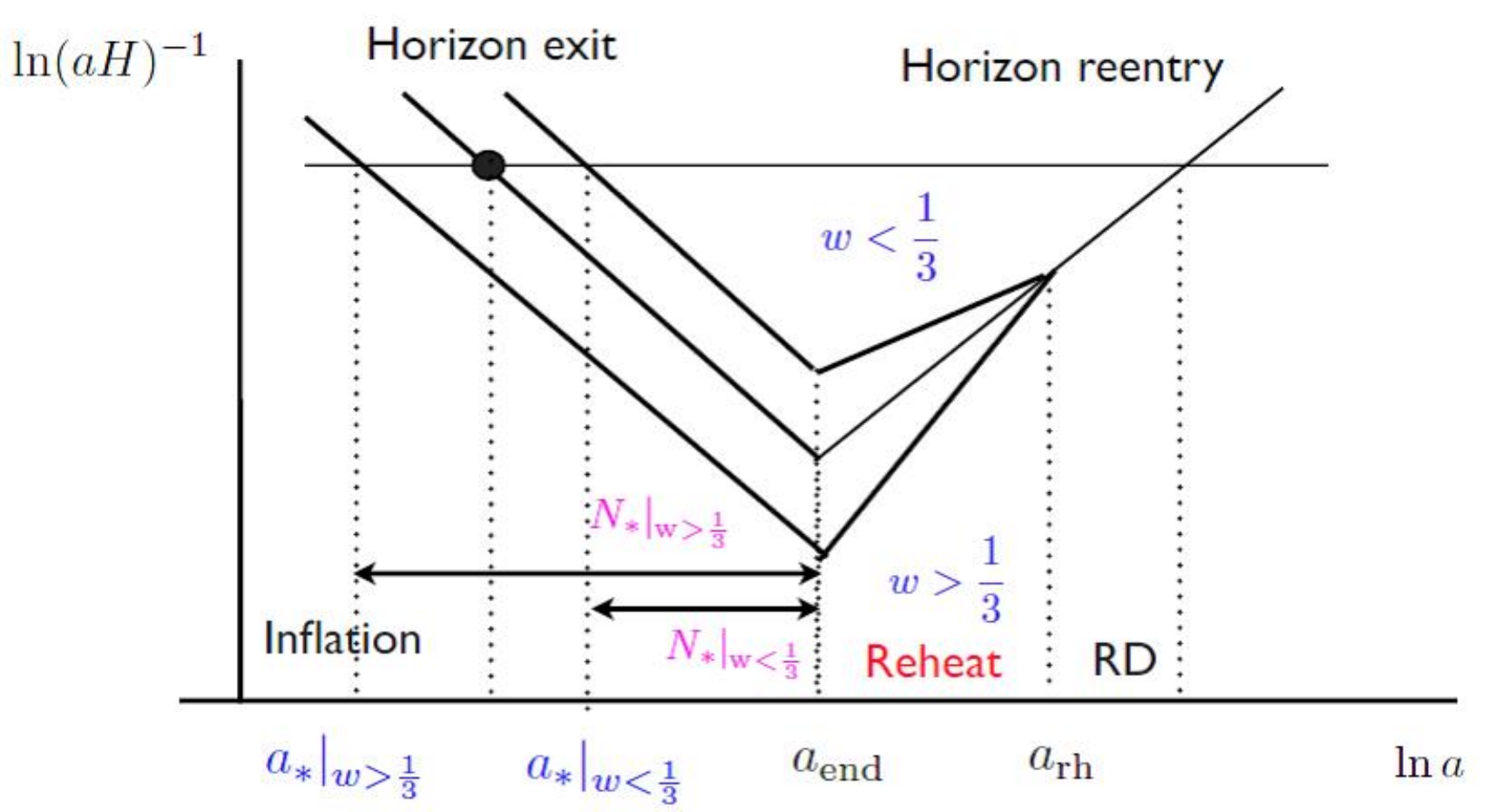}
    \end{center}
  \caption{Horizon distance as a function of scale factor (in log scales). The horizontal line corresponds to the scale of our interest entering the horizon during radiation.  The black dot is the moment that the scale of our interest exits the horizon. }
  \label{RH}
\end{figure}

If the reheating is delayed due to a late-time decay of the inflaton, depending on whether $w>\frac{1}{3}$ or not, the contribution coming from reheating dynamics can increase or decrease the number of efoldings during inflation. The reason is the following. For $w>\frac{1}{3}$, the horizon radius of the Universe at recombination would have been larger than in the case of instantaneous reheating. Therefore, the largest scale of cosmological interest must have left the horizon earlier so that the number of efoldings during inflation gets larger. On the other hand, for $w<\frac{1}{3}$, the situation is the opposite, meaning that the horizon radius of the Universe at recombination would have been smaller than in the case of instantaneous reheating and a smaller number of efoldings during inflation is needed. Lastly, for $w=\frac{1}{3}$, there is no distinction between reheating and radiation domination after reheating, so there is no effect from reheating dynamics on the number of efoldings in this case.  The details are depicted in Fig.~\ref{RH}.

 \section{Examples of inflection point inflation}
 
 In this section, we illustrate some examples for inflection point inflation in the context of the SM Higgs inflation and the $B-L$ Higgs inflation. Using the general results for the form of the effective potential near the inflection point in the previous section, we obtain the approximate formulas for the coefficients in each case and discuss the relation between the effective couplings and the fundamental parameters of the model.

 \subsection{SM Higgs inflation}
 
First we employ the general results for inflection point inflation to the case of the SM Higgs inflation minimally coupled to gravity and show that the number of efoldings in this case is insufficient. Then, we continue our discussion for the SM Higgs inflation with a non-minimal coupling to gravity and comment on the effects of reheating in this case.  

\subsubsection{The case without non-minimal coupling}

We discuss the SM Higgs inflation at the inflection point by using the general discussion in the previous section.
The RG-improved Higgs potential takes the following form,
\be
V(h)=\frac{1}{4}\lambda(h)h^4. \label{SMHiggs}
\ee 
where $\lambda(h)$ is the running quartic coupling evaluated at the Higgs field value \cite{sm2loops,thresholds}.
Here, near the criticality where the beta function for the effective Higgs quartic coupling vanishes, namely, $\beta=d\lambda/d\ln h=0$, 
the effective Higgs quartic coupling is given by
\be
\lambda(h)=\lambda_{\rm cr}+b \left(\ln \frac{h}{h_{\rm cr}}  \right)^2
\ee
where $b$ is the two-loop coefficient given by $b=0.4/(4\pi)^4$ and the Higgs field value at criticality is $h_{\rm cr}\sim 1$.

Now we expand the Higgs potential near the inflection point $h=h_0$ as follows,
\be
V(h)= V_0 + \lambda_1 (h-h_0)+\frac{1}{3!} \lambda_3 (h-h_0)^3
\ee
with
\bea
V_0 &=& \frac{1}{24}b \left(\frac{37}{12}-7\sqrt{\frac{25}{144}-\frac{\lambda_{\rm cr}}{b}} \right) h^4_0, \\
\lambda_1&=& \frac{2}{3}b \left(\frac{1}{3}- \sqrt{\frac{25}{144}-\frac{\lambda_{\rm cr}}{b}}\right) h^3_0, \\
\lambda_3&=& 6 b \sqrt{\frac{25}{144}-\frac{\lambda_{\rm cr}}{b}}\, h_0
\eea
where use is made of the inflection point condition $V^{\prime\prime}(h_0)=0$,
\be
\ln\frac{h_0}{h_{\rm cr}}=-\frac{7}{12}+ \sqrt{\frac{25}{144}-\frac{\lambda_{\rm cr}}{b}}.
\ee
Thus, the inflection point is located at $h_0<h_{\rm cr}$.
Requiring $\lambda_{\rm cr}<\frac{25}{144}b$ for real $\lambda_1, \lambda_3$, and $\lambda_{\rm cr}>\frac{b}{16}$ for positive $\lambda_1$, we can show that there is no false vacuum.  We also note that $V_0$ is always positive. 
Taking $\lambda_{\rm cr}\approx \frac{b}{16}\approx 10^{-6}$ for $\lambda_1\ll V_0$, we obtain the other parameters of the potential as
\bea
V_0&\approx& \frac{1}{32}b h^4_0=3.1\times 10^{-7}\left(\frac{b}{1.0\times 10^{-5}}\right) h^4_0,  \label{V0a} \\  
\lambda_3&\approx& 2 b h_0=2.0\times 10^{-5} \left(\frac{b}{1.0\times 10^{-5}}\right)h_0,  \label{l3a} \\
\lambda_1 &\approx& 3.4\times 10^{-7} \left(\frac{b}{1.0\times 10^{-5}}\right)^{3/2} h^6_0\, a^2, \label{l1a}
\eea
where $a$ depends on the tuning between $\lambda_{\rm cr}$ and $\frac{b}{16}$ and is matched to $a\approx \cos\Theta$ by the CMB normalization in eq.~(\ref{norm}),
and the inflection point as
\be
h_0 \approx e^{-\frac{1}{4}}\, h_{\rm cr}.  \label{inflectionpt}
\ee
In this case, however, as  the Planck bound on the tensor-to-scalar ratio leads to $h_0\,(b/1.0\times 10^{-5})^{1/4}\lesssim 0.32$, the number of efoldings during inflation is bounded by 
\be
N_{\rm max}\approx 0.53\,(1.0\times 10^{-5}/b)^{1/4}\,\frac{h_0^{1/2}}{a}.  \label{Nmaxa}
\ee
Taking $b=1.0\times 10^{-5}$ in the SM, the number of efoldings is too small.

\subsubsection{The case with non-minimal coupling}

We introduce a non-minimal coupling of the SM Higgs to gravity. 
The tree-level Jordan-frame action for the inflaton $h$ with non-minimal coupling is given by
\be
S=\int d^4 x\sqrt{-g}\left(\frac{1}{2}(1+\xi h^2) R-\frac{1}{2}(\partial_\mu h)^2-\frac{1}{4} \lambda h^4 \right)
\ee
where $\xi$ is the non-minimal gravity coupling.  
Then, by making a Weyl transformation of the metric with $g_{\mu\nu}= g^E_{\mu\nu}/\Omega^2$ with $\Omega=1+\xi h^2$, we obtain the Einstein-frame action as
\be
S= \int d^4x \sqrt{-g_E} \left(\frac{1}{2}R -\frac{1}{2}(\partial_\mu \chi)^2 -V_E  \right) 
\ee
where the canonical inflaton field is related to the Jordan frame Higgs by
\be
\frac{d\chi}{d h}= \frac{\sqrt{1+\xi (1+6\xi)h^2}}{1+\xi h^2}, \label{canchi}
\ee
and
 the inflaton potential in Einstein frame is given by
\be
V_E = \frac{1}{4} \lambda\varphi^4, \quad\quad \varphi\equiv  \frac{h}{\sqrt{1+\xi h^2}}. \label{SMHiggsn}
\ee

In the original Higgs inflation with tree level $\lambda$, the COBE normalization requires a large non-minimal coupling, $\xi\sim 10^4$ \cite{SMHiggs1}. Therefore, in this case, the unitarity scale in the vacuum is not far from the Hubble scale during inflation, making the semi-classical approximation of the Higgs inflation questionable \cite{unitarity1}. Nonetheless, it is known that the unitarity scale becomes background-field dependent so it becomes larger than the Hubble scale during inflation \cite{unitarity2}.  The Higgs inflation with non-minimal coupling is shown to be identical to the $R^2$ inflation or its scalar-tensor dual at the classical level \cite{R2inflation}, although there are distinctions at the loop level due to the would-be Goldstone bosons in the Higgs inflation case \cite{unitarity1}.  

At the loop-level, the effective inflaton quartic coupling can be identified from the one-loop Higgs potential with rescaled inflaton-dependent masses of particles and it turns out to be a function of $\varphi$ \cite{loops,sm2loops}.
Near the inflection point where the one-loop beta function vanishes, the loop corrections become more important and the effective inflaton quartic coupling is dominated by two-loop beta functions as follows,
\be
\lambda(\varphi)= \lambda_{\rm cr} + b \left(\ln \frac{\varphi}{\varphi_{\rm cr}} \right)^2.
\ee
We note that  for the SM Higgs boson, the two-loop beta function is given by $b=1.0\times 10^{-5}\equiv b_{\rm SM}$ \cite{sm2loops} whereas $\lambda_{\rm cr}$ and $\varphi_{\rm cr}$ depend on the precise values of top quark and gauge couplings in the low energy. 
The Higgs inflation near the inflection point was discussed in the literature \cite{Higgsinflection} and it was generalized to the singlet scalar inflation with an extra $U(1)$ gauge theory near the inflection point \cite{plateau}. In this work, we revisit the Higgs inflation near criticality (meaning a zero quartic coupling and its vanishing beta function) from the point of view of inflection point inflation. 

Now we are making an expansion of the effective inflaton potential around the inflection point $\chi=\chi_0$ for the canonical Higgs (equivalently to $\varphi=\varphi_0$) as follows,
\be
V_E = V_0 + \lambda_1 (\chi-\chi_0)+ \frac{1}{3!} \lambda_3 (\chi-\chi_0)^3
\ee
where
\bea
V_0&=&\frac{1}{4}\left(\lambda_{\rm cr} + b \ln^2 \frac{\varphi_0}{\varphi_{\rm cr}} \right) \varphi^4_0,  \label{v00}\\ 
\lambda_1 &=& \frac{\partial V_E}{\partial \chi}\bigg|_{\chi_0} = \frac{d h}{d \chi}\frac{d\varphi}{d h} \frac{d V_E}{d\varphi}\bigg|_{\chi_0}, \label{lam11}\\
\lambda_3 &=& \frac{\partial^3 V_E}{\partial \chi^3}\bigg|_{\chi_0}=\frac{dh}{d\chi} \frac{\partial }{\partial h}\left(\frac{d^2 V_E}{d\chi^2} \right)\bigg|_{\chi_0}.  \label{lam33}
\eea
Here, the inflection point is determined from
\be
0=\frac{\partial^2 V_E}{\partial \chi^2}\bigg|_{\chi_0}=\frac{dh}{d\chi}\frac{\partial }{\partial h}\left(\frac{\partial V_E}{\partial \chi} \right)\bigg|_{\chi_0}.
\ee

Henceforth, we assume that $\xi\gtrsim 1$.
Then, from eq.~(\ref{canchi}), taking $\xi h^2\gtrsim 1$ during inflation, the canonical field is related to the inflaton field in Jordan frame by $\chi\simeq\sqrt{6}\ln (h/\sqrt{\xi})$, resulting in $\frac{dh}{d\chi}\simeq\frac{h}{\sqrt{6}}$. In this case, we also get $\varphi\simeq\frac{1}{\sqrt{\xi}}\Big(1-\frac{1}{2\xi h^2}\Big)$, resulting in $\frac{d\varphi}{dh}\simeq\frac{1}{\xi^{3/2}h^3}$.
Therefore, we get the derivatives of the potential as follows,
\bea
\frac{\partial V_E}{\partial \chi} &\simeq& \frac{1}{\sqrt{6}\xi^{3/2}h^2}\frac{\partial V_E}{\partial\varphi}, \\
\frac{\partial^2 V_E}{\partial \chi^2} &\simeq& -\frac{1}{3\xi^{3/2}h^2}\frac{\partial V_E}{\partial\varphi}+\frac{1}{6\xi^3 h^4} \frac{\partial^2 V_E}{\partial \varphi^2}, \\
\frac{\partial^3 V_E}{\partial \chi^3}&\simeq & \frac{2}{3\sqrt{6}\xi^{3/2}h^2} \frac{\partial V_E}{\partial\varphi}-\frac{1}{\sqrt{6}\xi^3 h^4}\frac{\partial^2 V_E}{\partial\varphi^2}+\frac{1}{6\sqrt{6}\xi^{9/2}h^6}\frac{\partial^3 V_E}{\partial\varphi^3}.
\eea
Consequently, from eqs.~(\ref{v00})-(\ref{lam33}), the model parameters can be computed at the inflection point $\chi_0$ or $\varphi_0\simeq 1/\sqrt{\xi}$ as
\bea
V_0 & \simeq & \frac{1}{4\xi^2}\left(\lambda_{\rm cr} + b \ln^2 \frac{1}{\varphi_{\rm cr}\sqrt{\xi}}\right), \label{nV0}\\
\lambda_1 &\simeq & \frac{1}{\sqrt{6}\xi^{3/2}h^2}\frac{\partial V_E}{\partial \varphi}\bigg|_{\varphi_0}, \label{nl1}\\
\lambda_3 &\simeq & -\frac{4}{3\sqrt{6}\xi^{3/2}h^2}\frac{\partial V_E}{\partial\varphi}\bigg|_{\varphi_0}+\frac{1}{6\sqrt{6} \xi^{9/2}h^6} \frac{\partial^3 V_E}{\partial\varphi^3}\bigg|_{\varphi_0}  \label{nl3}
\eea 
where we imposed the inflection point condition,
\be
\frac{\partial V_E}{\partial \varphi}\bigg|_{\varphi_0}= \frac{1}{2\xi^{3/2}h^2}\frac{\partial^2 V_E}{\partial \varphi^2}\bigg|_{\varphi_0}.
\ee

Imposing $\frac{\partial^2 V_E}{\partial\varphi^2}\Big|_{\varphi_0}$ to be small, we obtain $\lambda_{\rm cr}\simeq \frac{b}{16}$ and $\varphi_0 \simeq e^{-\frac{1}{4}}\varphi_{\rm cr}\simeq 1/\sqrt{\xi}$, leading to the non-minimal coupling, $\xi\simeq e^{\frac{1}{2}}\varphi^{-2}_{\rm cr}$. 
Then, for $h_0=\frac{c}{\sqrt{\xi}}$ with $c\gtrsim 1$, the effective couplings near the inflection point are determined as
\bea
V_0 &\simeq&  \frac{b}{32\xi^2}\simeq (3.1 \times10^{-7}) \left(\frac{b}{1.0\times 10^{-5}}\right)\frac{1}{\xi^2}, \\
\lambda_3 &\simeq & \frac{b}{3\sqrt{6} c^6 \xi^2}\simeq (1.4\times 10^{-6})  \left(\frac{b}{1.0\times 10^{-5}}\right)\frac{1}{c^6 \xi^2}, \\
\lambda_1&\simeq& (3.2\times 10^{-7}) \left(\frac{b}{1.0\times 10^{-5}}\right)^{3/2} \frac{a^2}{\xi^3}.\label{lam1}
\eea
Here, we note that $a$ depends on the tuning between $\lambda_{\rm cr}$ and $\frac{b}{16}$, determined by the low-energy parameters in the RG equation for the quartic coupling, and it is matched to $a\approx \cos\Theta$ by the CMB normalization (\ref{norm}) for given ratios of inflaton potential parameters and $N_*$ from eq.~(\ref{Nstar}).
As compared to the case with minimally coupled Higgs boson in eqs.~(\ref{V0a})-(\ref{l1a}), the effective couplings of the inflection point inflation can be suppressed for large $\xi$ and/or $c$, meaning that the range of the approximate inflection point is extended.

\begin{figure}
  \begin{center}
\includegraphics[height=0.40\textwidth]{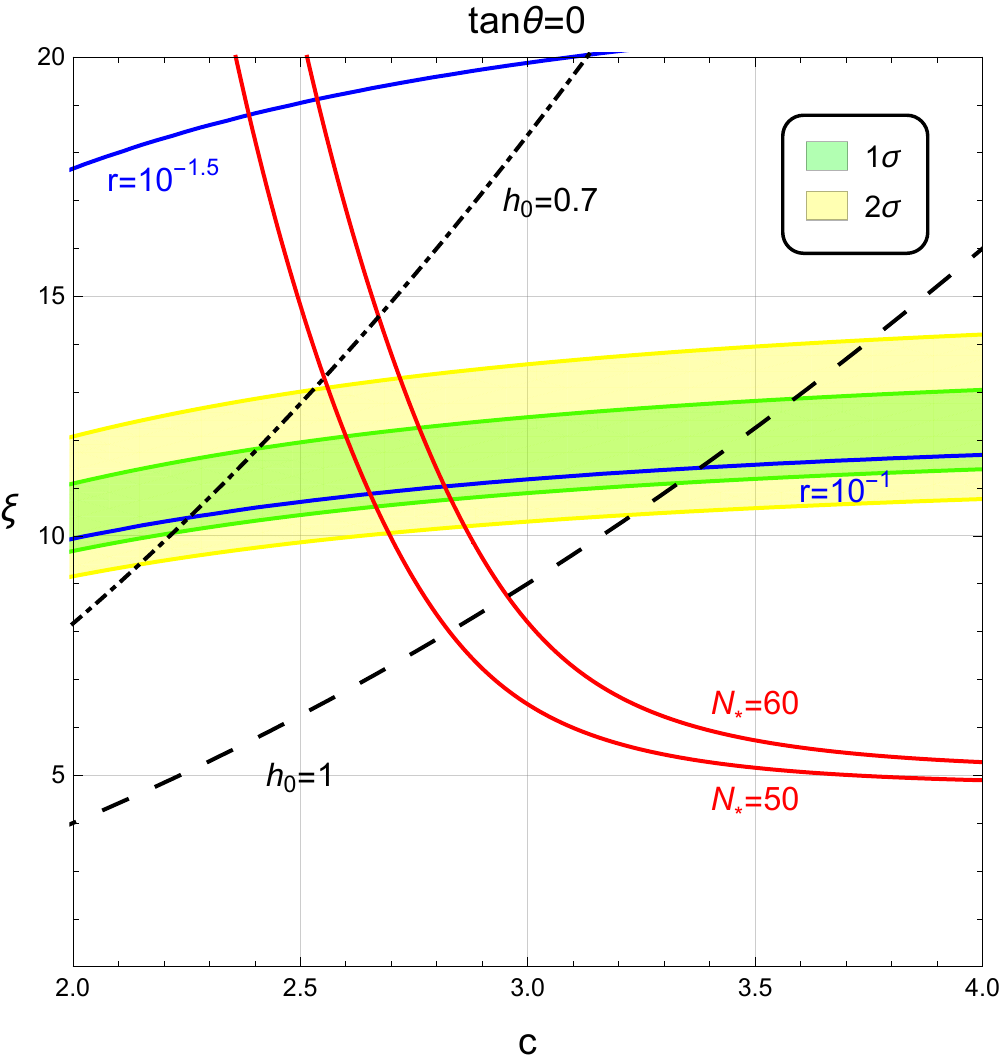}
   \includegraphics[height=0.40\textwidth]{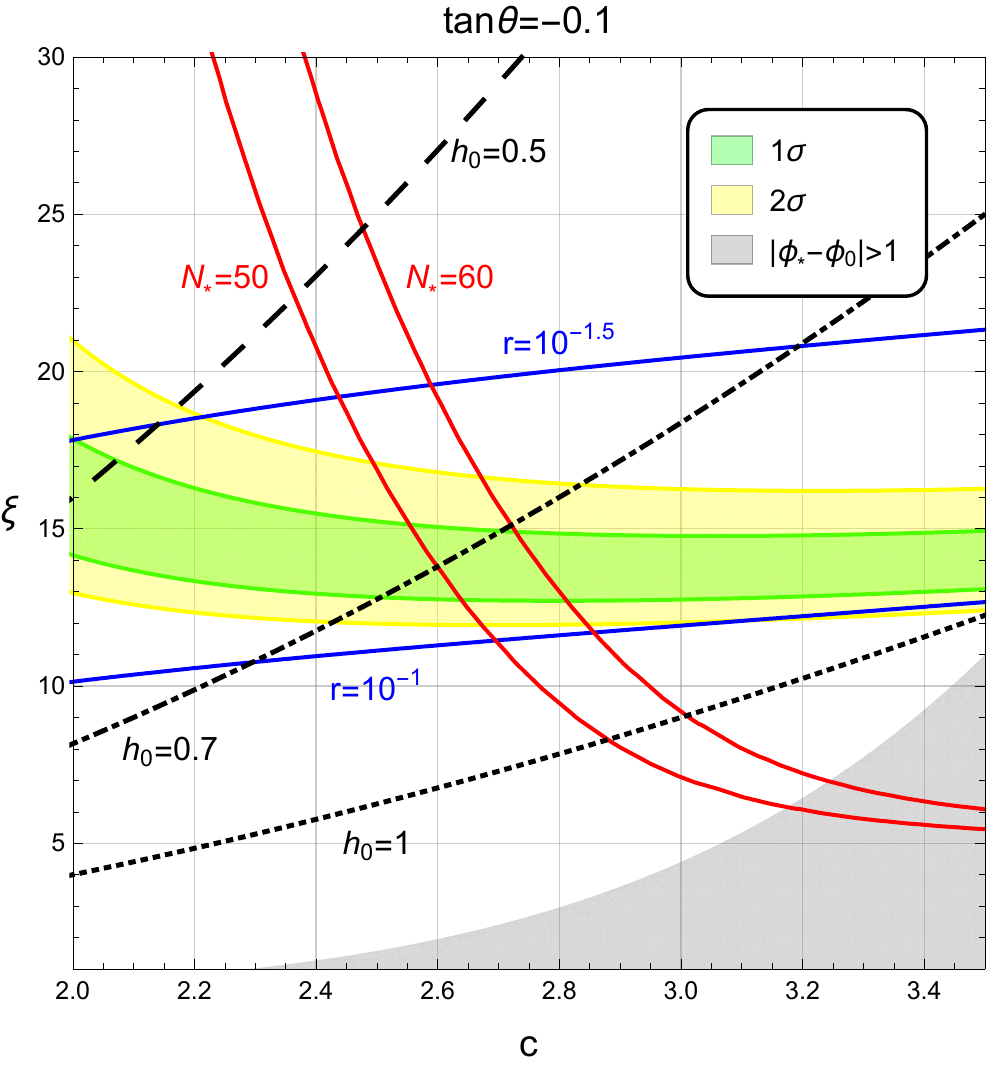} \\
   \includegraphics[height=0.40\textwidth]{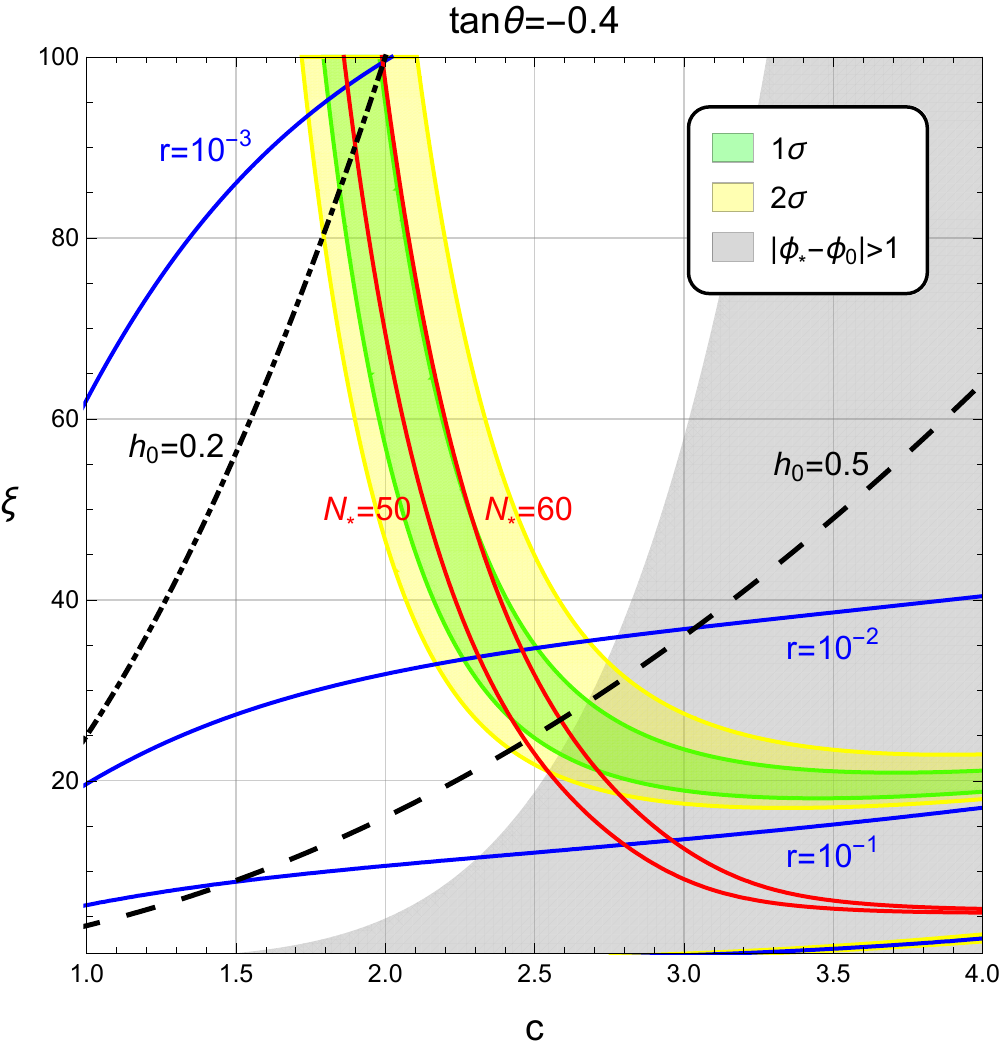}
   \includegraphics[height=0.40\textwidth]{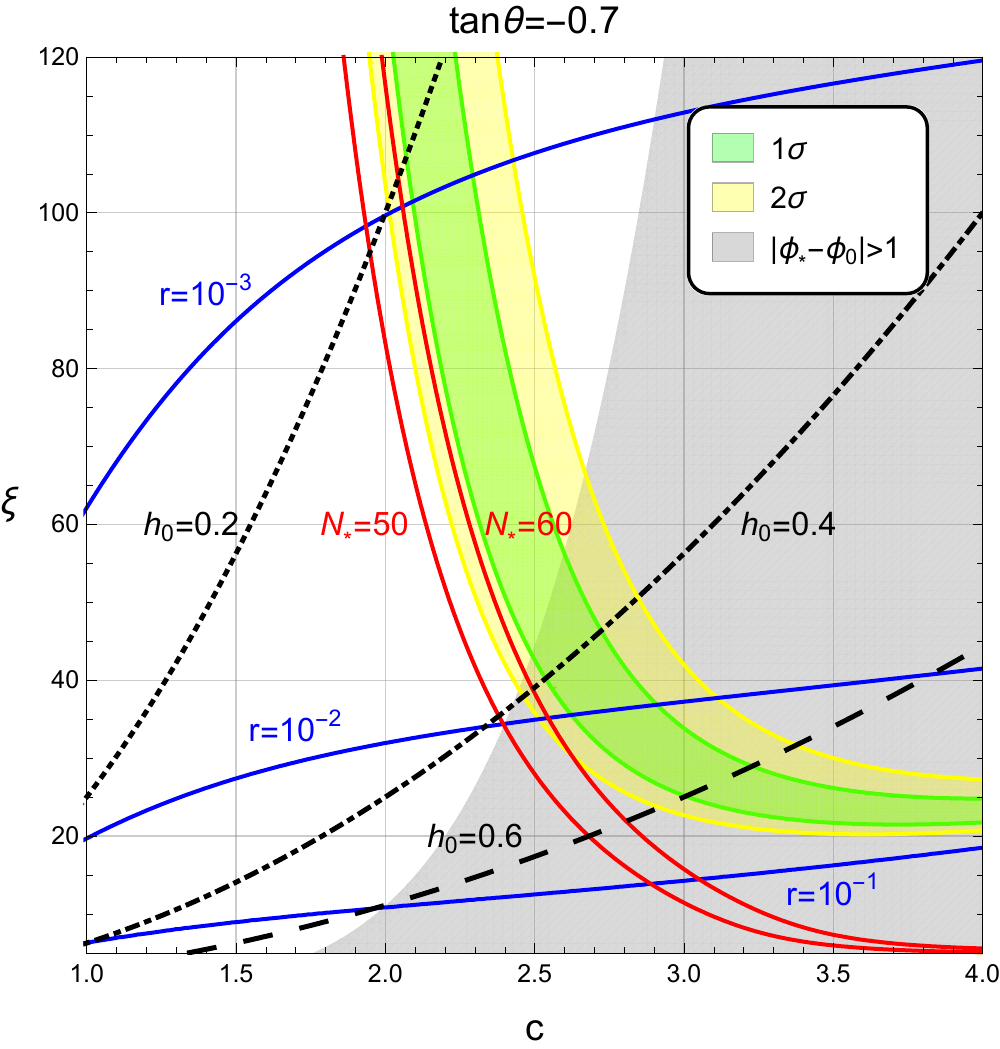}
   \end{center}
  \caption{Parameter space for the inflection point $c=h_0\sqrt{\xi}$ and the non-minimal coupling $\xi$ for $\tan\Theta=0, -0.1,-0.4,-0.7 $ in red lines with $N_*=50, 60$. The regions favored by the spectral index from Planck within $1\sigma$ and $2\sigma$ are shown in green and yellow, respectively. Tensor-to-scalar ratio with $r=10^{-1}, 10^{-1.5} $ or $r=10^{-1}, 10^{-2}, 10^{-3}$ are shown in blue lines in the upper and lower panels, respectively. The region with inflaton field value taking $|\phi_*-\phi_0|>1$ at horizon exit is shown in gray. Severa values of $h_0$ are shown in black lines too.  Two-loop beta function coefficient is chosen to $b=1.0\times 10^{-5}$. }
  \label{Higgs-parameter}
\end{figure}

We have checked that the higher order terms in the inflaton potential near the inflection point are suppressed by a sizable $\xi$ or a large inflation field value in the relevant parameter space.
Therefore, the number of efoldings during inflation is bounded by 
\be
N_{\rm max}\simeq 2.1\,(1.0\times 10^{-5}/b)^{1/4}\,\frac{\xi^{1/2} c^3}{a},  \label{Nmax}
\ee
which is enhanced for a large non-minimal coupling and/or a large inflation field value at the inflection point, as compared to the case with minimally coupled Higgs boson in eq.~(\ref{Nmaxa}). 
From the Planck bound on the tensor-to-scalar ratio, we get the upper bound on the non-minimal coupling as $\xi\gtrsim 9.6\,(b/1.0\times 10^{-5})^{1/2}$. Consequently, a relatively small non-minimal coupling is allowed in inflection point inflection as compared to the classical Higgs inflation. 
In this case, the field-dependent cutoff given by $M_P/\sqrt{\xi}$ is much larger than the Hubble scale of order $\sqrt{\lambda(\varphi_I)} M_P/\xi$, which is suppressed for a small $\lambda(\varphi_I)$ for inflation field values $\varphi_I$.

Imposing $N_{\rm max}\sim 120$ in eq.~(\ref{Nmax}) with the bound on tensor-to-scalar ratio, $\xi\gtrsim 9.6\,(b/1.0\times 10^{-5})^{1/2}$, we get the upper bound on the inflaton field value at the inflection point as $c\lesssim 2.6 \,a^{1/3}$, independent of the value of $b$.  On the other hand, using $\frac{V_0}{\lambda_3}\simeq 0.2 c^6$ from eqs.~(\ref{nV0}) and (\ref{nl3}), we find a trans-Planckian field excursion of the canonical inflation field  during inflation for $c\gtrsim 1.3$.  
Therefore, in the SM Higgs inflation with non-minimal coupling, the inflation field values near the inflection point are necessarily trans-Planckian for $a={\cal O}(1)$ in terms of the canonical inflation field $\chi$.
Nonetheless, the inflaton field in the Jordan frame can be sub-Planckian, because $h_0=c/\sqrt{\xi}\simeq (2.1 c^4/N_{\rm max})(1.0\times 10^{-5}/b)^{1/4}/a\lesssim 1$ for $b>b_{\rm SM}$ and/or a larger $\xi$ from the bound on the tensor-to-scalar ratio.

\begin{figure}
  \begin{center}
\includegraphics[height=0.40\textwidth]{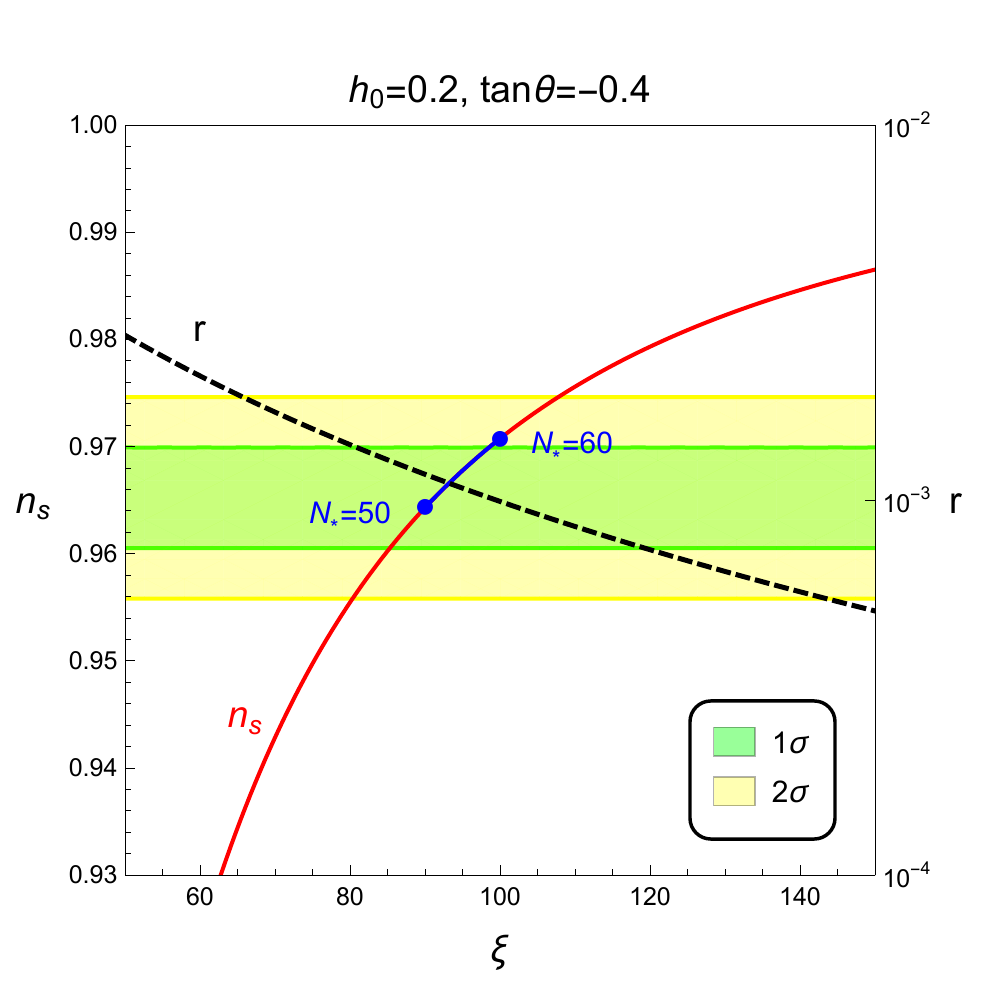}
   \includegraphics[height=0.40\textwidth]{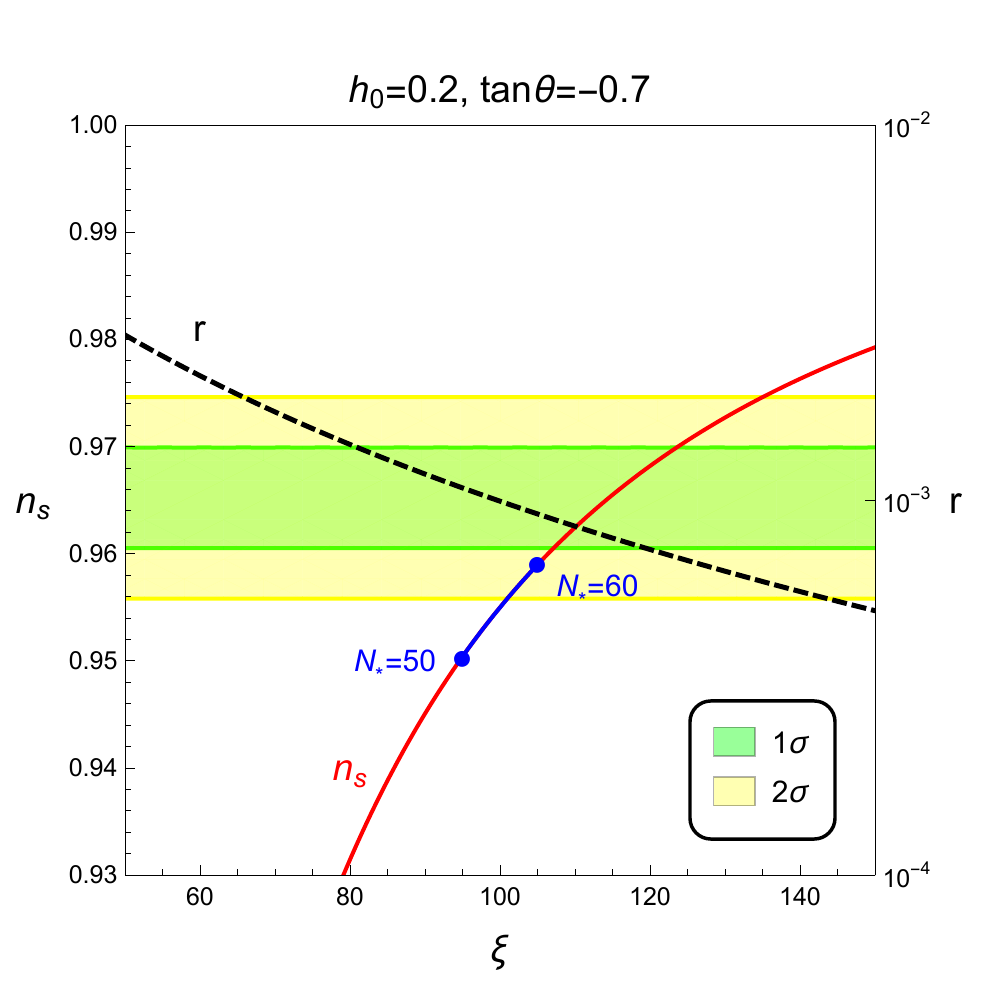} \\
   \includegraphics[height=0.40\textwidth]{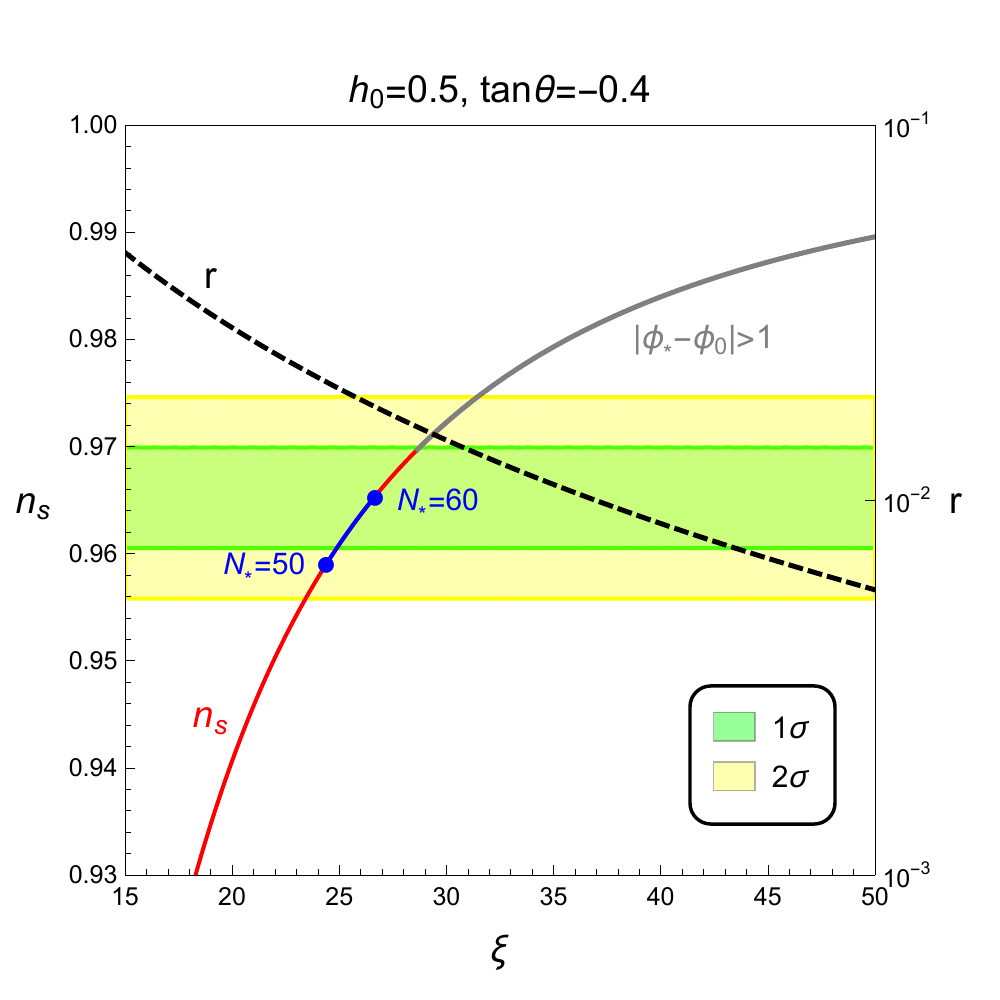}
   \includegraphics[height=0.40\textwidth]{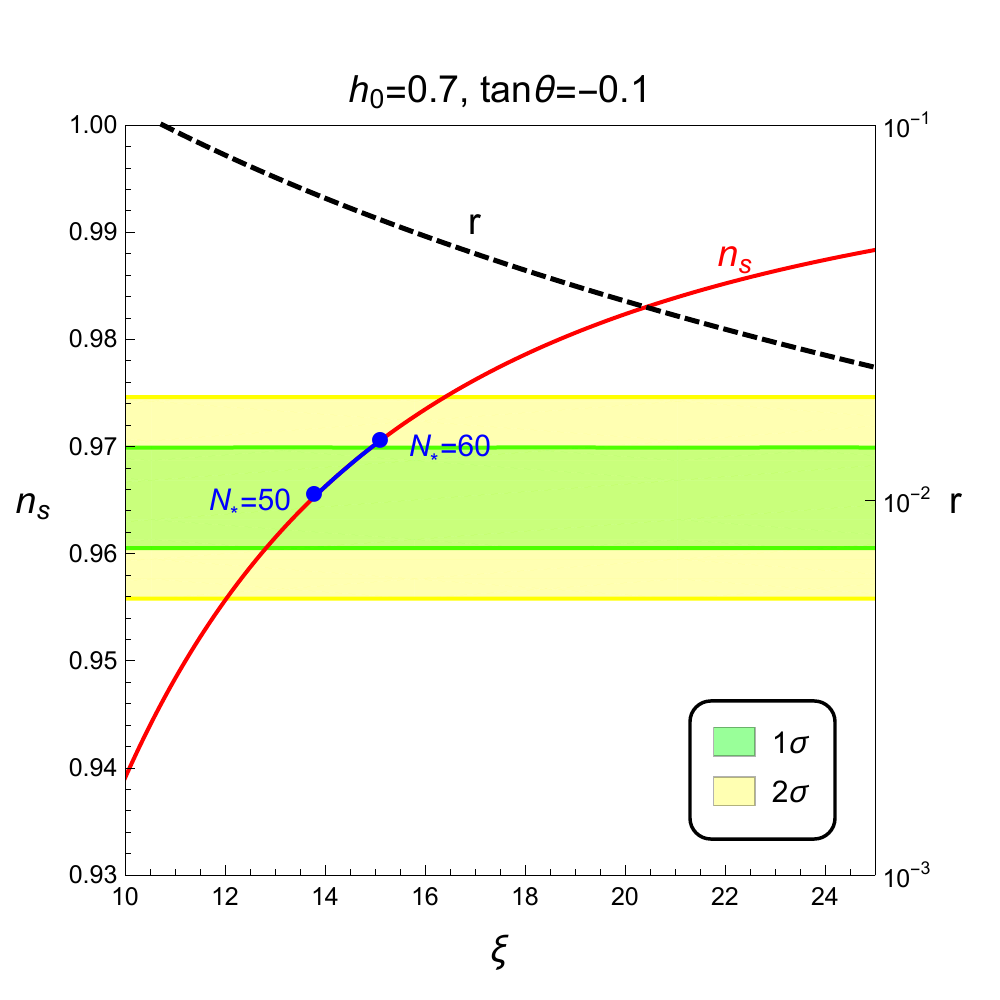}
      \end{center}
  \caption{The spectral index as a function of the non-minimal coupling $\xi$ in the SM Higgs inflation in red lines, for a set of inflection point $h_0$ and $\tan\Theta$, $(0.2,-0.4), (0.2, -0.7), (0.7, -0.1), (0.5,-0.4)$, clockwise. Blue lines correspond to the spectral index for the number of efoldings between $N_*=50$ and $60$.  The regions favored by the spectral index from Planck within $1\sigma$ and $2\sigma$ are shown in green and yellow, respectively. The tensor-to-scalar ratio is shown in black line  in each plot. The Planck $1\sigma, 2\sigma$ bands of the spectral index are shown too. The region with inflaton field value taking $|\phi_*-\phi_0|>1$ at horizon exit is shown in gray line. Two-loop beta function coefficient is chosen to $b=1.0\times 10^{-5}$. 
}
  \label{Higgs-ns}
\end{figure}

In Fig.~\ref{Higgs-parameter}, employing the approximate potential up to cubic terms near the inflection point and taking the two-loop beta function coefficient to $b=1.0\times 10^{-5}$, we show the parameter space for $\xi$ and $c$ for a chosen $\tan\Theta=0, -0.1,-0.4,-0.7 $ in red lines with $N_*=50, 60$ while the spectral index favored by Planck $1\sigma$ and $2\sigma$ bounds is shown in green and yellow regions, respectively. 
Furthermore, in the same figures, the lines with tensor-to-scalar ratio, $r=10^{-1}, 10^{-1.5}$
and  $r=10^{-1}, 10^{-2}, 10^{-3}$ are shown in blue lines, in upper and lower panels, respectively, and  the region with inflation field value given by $|\phi_*-\phi_0|>1$ at horizon exit is shown in gray too. Several values of $h_0$ are shown along the black lines in each plot, thus telling us that the Higgs field value at inflection point in Jordan frame is sub-Planckian in most of the parameter space.  
For a chosen value of $\tan\Theta$, the $\lambda_1$ coupling is accordingly fixed by CMB normalization. 

In the case with a small $\tan\Theta$ in the upper panel, the measured spectral index favors the non-minimal coupling $\xi$ between $10$ and $13$  on left ($13$ and $17$ on right) and the inflection point $c$ between $2.5$ and $2.9$ in both cases.  Moreover, in this case, the tensor-to-scalar ratio tends to be large, saturating the upper bound from Planck. 

On the other hand, in the case with a large $\tan\Theta$ in the lower panel, where the region with $|\phi_*-\phi_0|>1$ at horizon exit gets larger, the favored region is given by the non-minimal coupling $\xi$ larger than $20$($30$) and the inflection point $c$ smaller than $2.8$($2.6$) on left (right). Moreover, in this case, the tensor-to-scalar ratio becomes smaller than the Planck bound in most of the parameter space favored by the measured spectral index.
We note that $|\tan\Theta|$ is favored by the spectral index to be smaller than about $0.7$.

In Fig.~\ref{Higgs-ns}, we also depict the spectral index as a function of the non-minimal coupling in red lines, for several choices of inflection point $h_0$ and $\tan\Theta$ to $(0.2,-0.4), (0.2, -0.7)$, $(0.7, -0.1), (0.5,-0.4)$, in the plots clockwise.  In each plot, the spectral index is bounded to blue line for the number of efoldings between $N_*=50$ and $60$ and  the spectral index favored by Planck $1\sigma$ and $2\sigma$ bounds is shown in green and yellow regions, respectively.   Two-loop beta function coefficient  is chosen to $b=1.0\times 10^{-5}$. We show the obtained tensor-to-scalar ratio in black line in the same figures as well.  The region with inflation field value taking $|\phi_*-\phi_0|>1$ at horizon exit is shown in gray line in the case with $(0.5,-0.4)$. 

We note that in the region with a relatively small non-minimal coupling in the lower panel of Fig.~\ref{Higgs-ns}, satisfying the observed spectral index, the tensor-to-scalar ratio tends to be large and sensitive to the choice of the inflection point. The larger the value of $h_0$, the smaller value of $\xi$ the observed spectral index is obtained for, and the larger the tensor-to-scalar ratio.

Now we remark on the dependence of the inflationary predictions on the reheating dynamics. 
The SM Higgs couples to the rest of the SM particles strongly, so the reheating temperature is high and expected to be of similar order to the one for the classical SM Higgs inflation, which is $3-15\times 10^{13}\,{\rm GeV}$ \cite{Higgsreheat}. Therefore, the number of efoldings during reheating could not be sizable, so the inflationary predictions are little dependent on the reheating dynamics.
In this case, the inflationary predictions with instantaneous reheating or $w=\frac{1}{3}$ are reliable.

\subsection{$B-L$ Higgs inflation}

The $U(1)_{B-L}$ symmetry is an anomaly-free gauge theory that is introduced as a natural extension of the SM with three Right-Handed(RH) neutrinos for generating neutrino masses via see-saw mechanism \cite{chun}. 
In the $U(1)_{B-L}$ extension of the SM,  we consider the effective potential for the $B-L$ Higgs $\phi$ which carries $+2$ charge under the $U(1)_{B-L}$ and has a nonzero non-minimal coupling to gravity. Then, we identify the $B-L$ Higgs field as the inflaton.
The Coleman-Weinberg inflation in the context of the $B-L$ Higgs inflation was discussed \cite{CW} 
and the $B-L$ running inflation with non-minimal coupling was also recently studied \cite{CWxi}.  

Near the critical point $\phi=\phi_{\rm cr}$, we choose the $B-L$ Higgs quartic coupling to $\lambda_\phi(\phi_{\rm cr})\approx 0$ and one-loop beta function to $\beta^{(1)}_{\lambda_{\phi}}(\phi_{\rm cr})\approx 0$.  In this case, we need to choose
the quartic coupling $\lambda_\phi$ for the $B-L$ Higgs to be small at low energies as for the SM Higgs. 
Then, ignoring  $\lambda_\phi$ in eq.~(\ref{lamphi1}) in Appendix A, we can obtain the relation between the RH neutrino Yukawa  coupling  $y_N$ and the $B-L$ gauge coupling $g_{B-L}$ at the critical point as
\bea
y_N\approx 2^{\frac{5}{4}} g_{B-L}.  \label{critical}
\eea
As a consequence,  the $B-L$ gauge boson and the RH neutrinos receive masses of similar order from the VEV of the $B-L$ Higgs, because $m_{Z'}=2 g_{B-L} \langle\phi\rangle$ and $m_N=\frac{1}{\sqrt{2}}\, y_N \langle\phi\rangle$ for a canonical scalar $\phi$.

After the condition (\ref{critical}) at the critical point is imposed, the two-loop beta function for the $B-L$ Higgs quartic coupling, eq.~(\ref{lamphi2}),
becomes
\bea
\beta^{(2)}_{\lambda_\phi}\approx \frac{14620 g^6_{B-L}}{(4\pi)^4}\equiv b_{\rm B-L}. 
\eea
For $g_{B-L}=0.16$, we get $b_{B-L}=10^{-5}$, which is of similar order to the SM value. 
Depending on the low energy value of $\lambda_\phi$, the critical point $\phi=\phi_{\rm cr}$ varies. 
For a fixed $\lambda_\phi$, one can determine $g_{B-L}$ and $y_N$ at low energies under the condition of the critical point in the UV, resulting in a definite prediction for  the ratio of masses of $Z'$ gauge boson and RH neutrinos.

Since $\lambda_\phi$ is small at low energies, the $B-L$ Higgs tends to be light \footnote{The $B-L$ symmetry can be broken with a tree-level negative mass squared for the $B-L$ Higgs or by the Coleman-Weinberg mechanism \cite{chun}. In the latter case, the $B-L$ Higgs without non-minimal coupling was discussed in view of Planck data as an inflaton candidate \cite{CW}. } so it decays dominantly into a pair of the SM Higgs bosons, reheating the SM sector.  Then,  the reheating temperature is given by
\bea
T_{\rm rh}= 0.5 \sqrt{M_P \Gamma_\phi}\approx 0.05\lambda_{H\phi} v_{B-L} \sqrt{\frac{M_P}{m_\phi}}
\eea
where use is made of $\Gamma_\phi\approx \frac{\lambda^2_{H\phi} v^2_{B-L}}{32\pi m_\phi}$ for $m_\phi\gg m_h$ with $m_h$ being the SM Higgs mass.  Thus, imposing the bound on the mixing angle between the $B-L$ Higgs and the SM Higgs from the Higgs signal strength, such as $|\sin\theta| \approx \frac{|\lambda_{H\phi}| v_{B-L} v}{m^2_\phi}<0.44$ \cite{Higgsfit}, we obtain the lower bound on the reheating temperature as
\bea
T_{\rm rh} \lesssim (5.3\times 10^8\,{\rm GeV}) \Big(\frac{m_\phi}{v} \Big)^{3/2}.
\eea

Suppose that the Higgs-portal coupling $\lambda_{H\phi}$ is very small in the UV so it is generated radiatively in the presence of the gauge kinetic mixing \cite{chun}. In this case, when the Higgs mass parameter is solely from the Higgs-portal coupling after the $B-L$ breaking, we get $|\lambda_{h\phi}|=\frac{m^2_h}{v^2_{B-L}}$ \cite{chun}, which satisfies the limit from the Higgs signal strength for $v_{B-L}\gg m_h$. Therefore,  the reheating temperature  can be much smaller as follows,
\bea
T_{\rm rh}\approx (3.2 \times 10^8\,{\rm GeV}) \Big(\frac{v^3}{v^2_{B-L} m_\phi} \Big)^{1/2} \ll 3.2 \times 10^8\,{\rm GeV}.
\eea
Therefore, depending on the equation of state during reheating, the reheating dynamics can affect much the inflationary predictions of the $B-L$ Higgs inflation.

\begin{figure}
  \begin{center}
\includegraphics[height=0.30\textwidth]{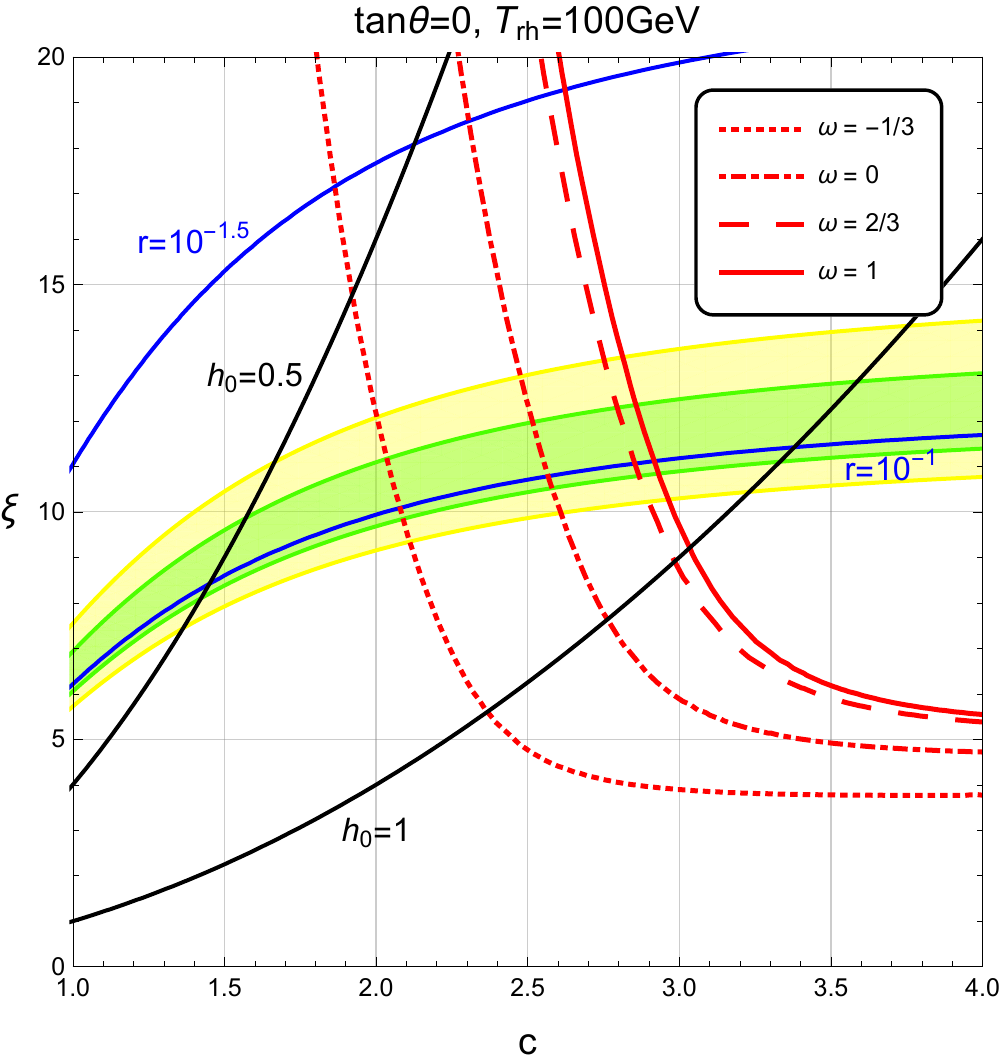}
   \includegraphics[height=0.30\textwidth]{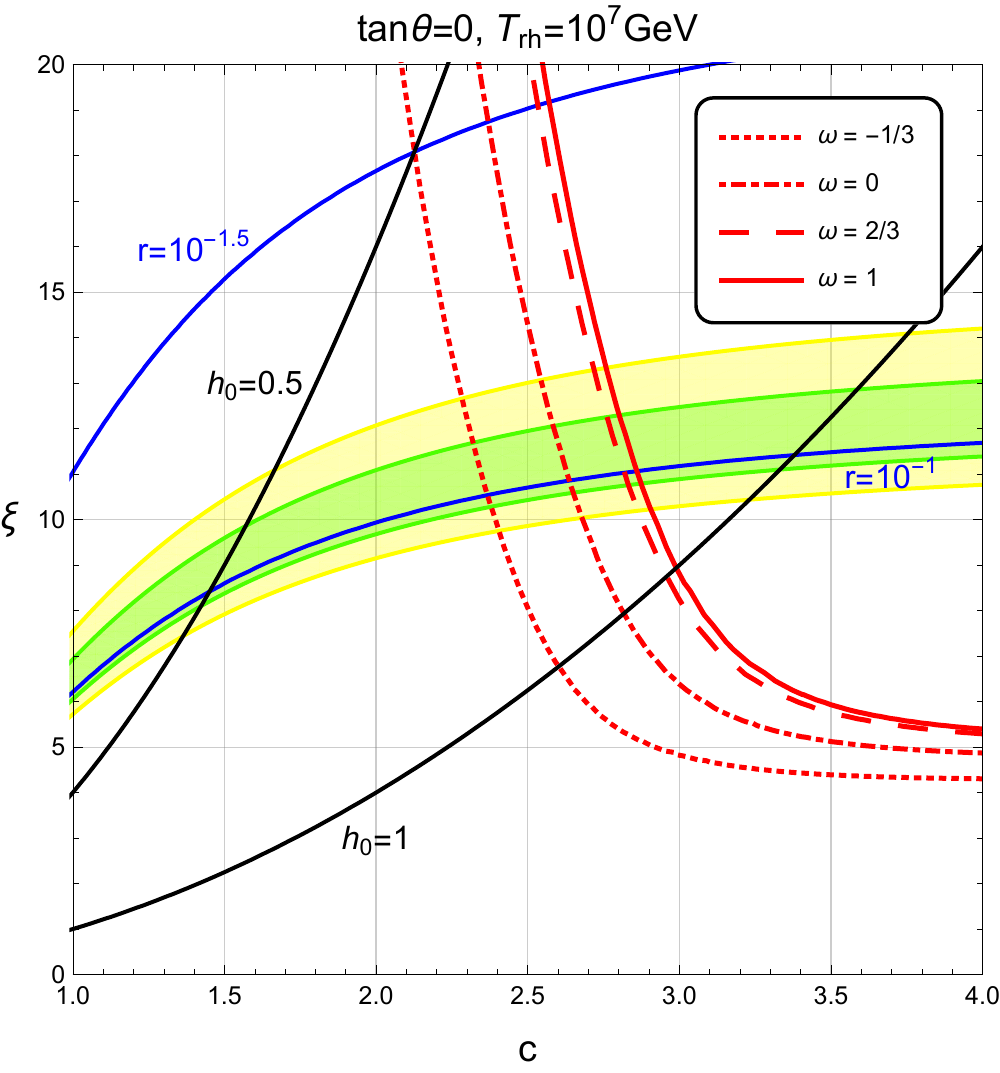}
      \includegraphics[height=0.30\textwidth]{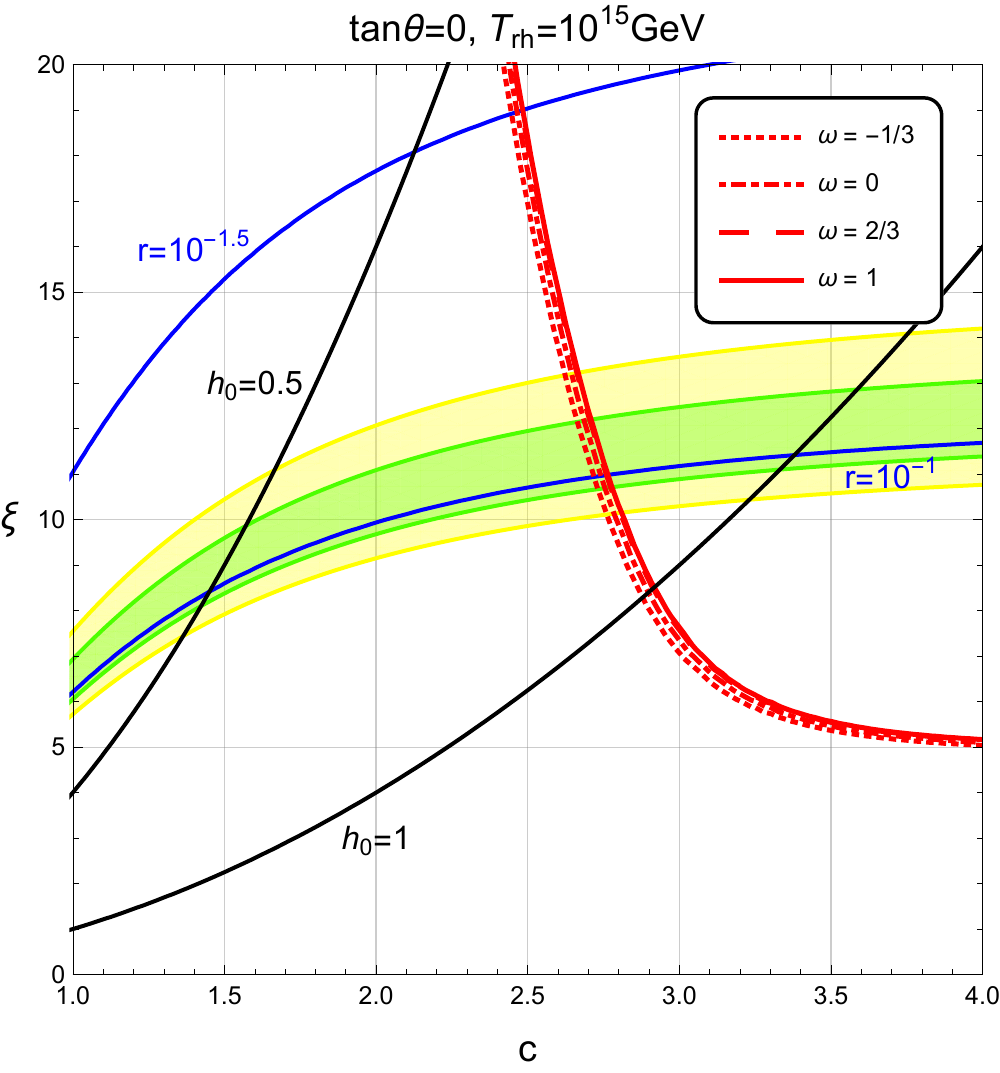} \\
   \includegraphics[height=0.30\textwidth]{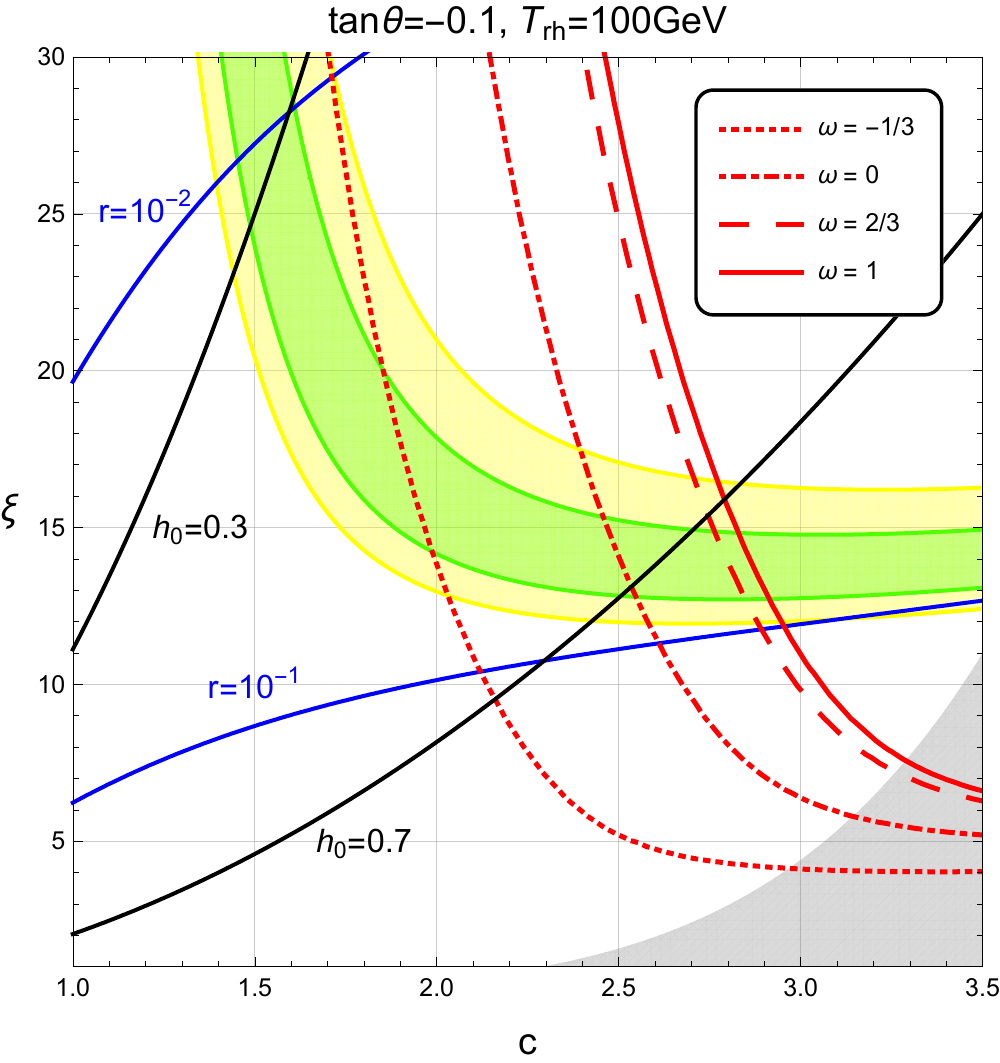}
    \includegraphics[height=0.30\textwidth]{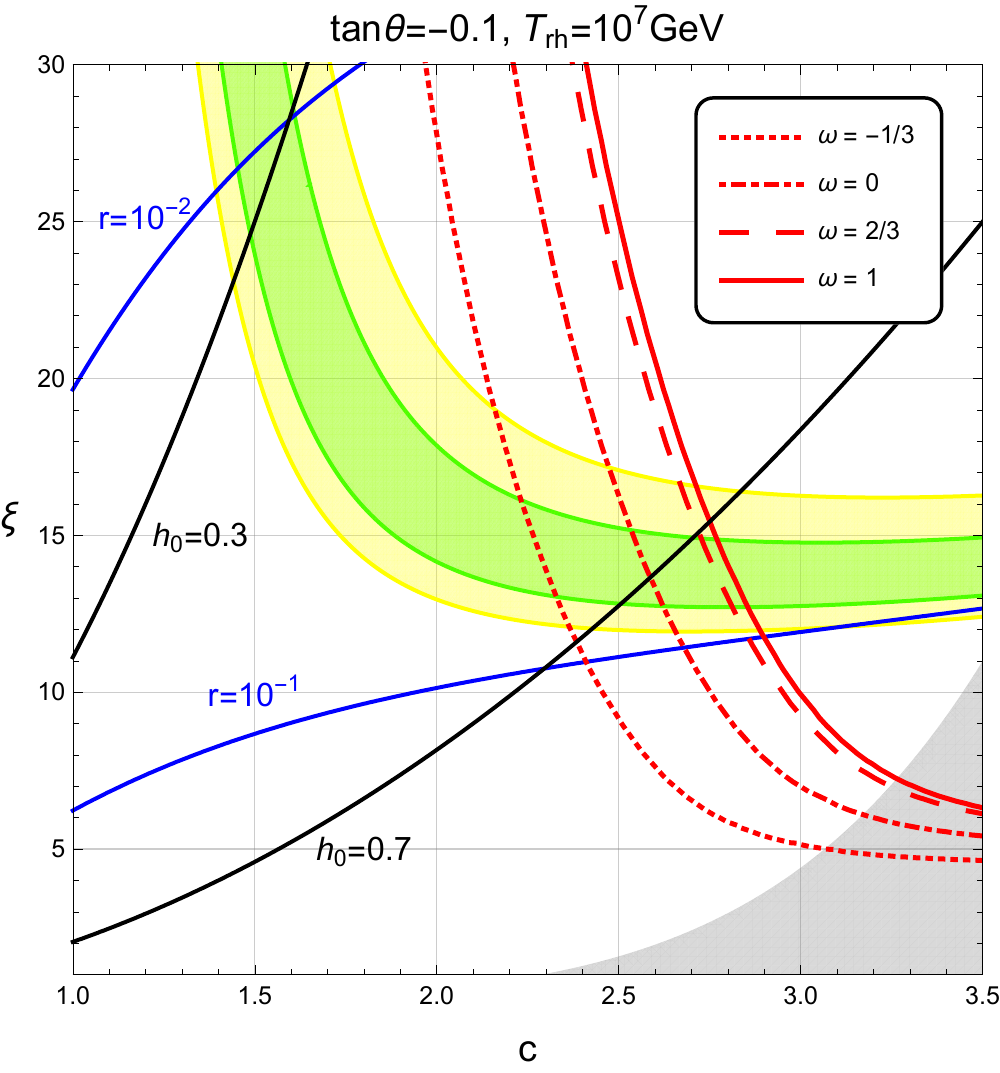}
   \includegraphics[height=0.30\textwidth]{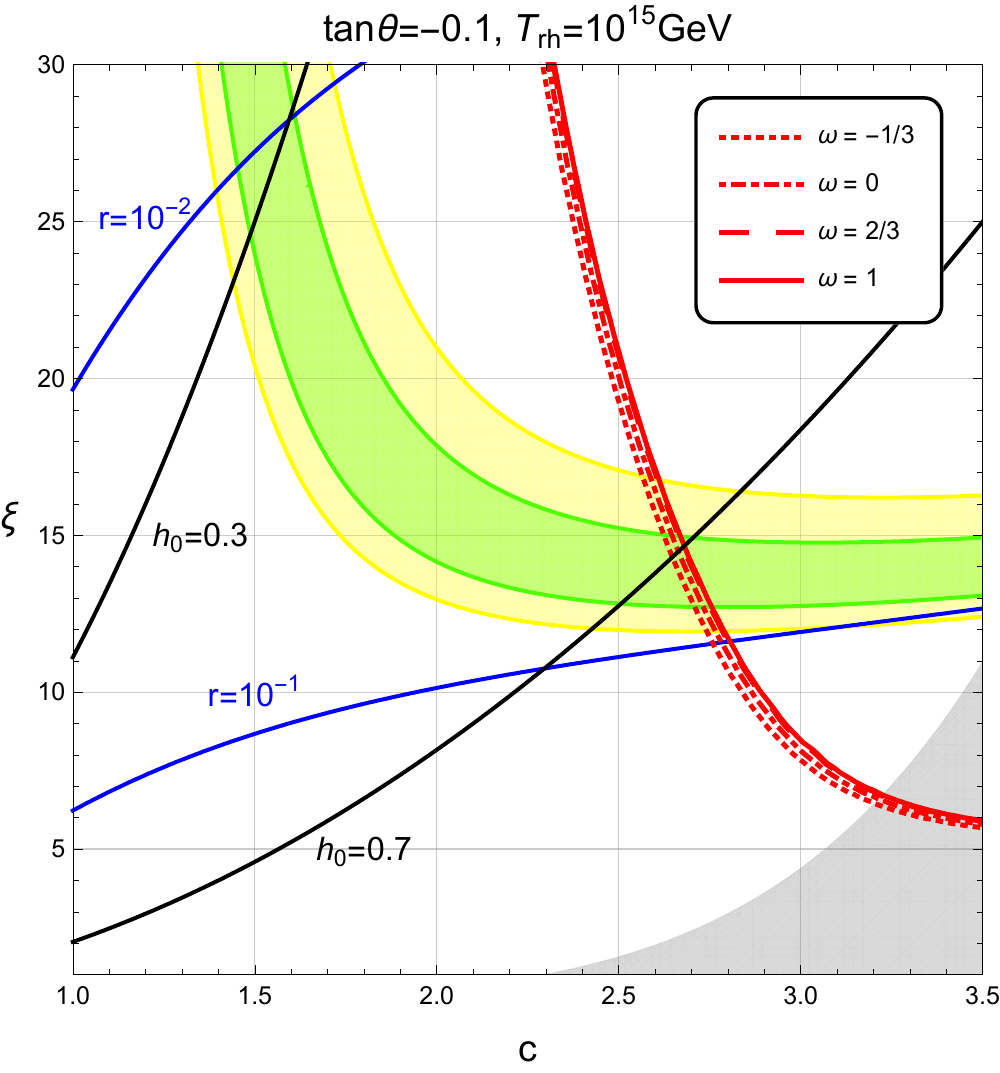}\\
   \includegraphics[height=0.30\textwidth]{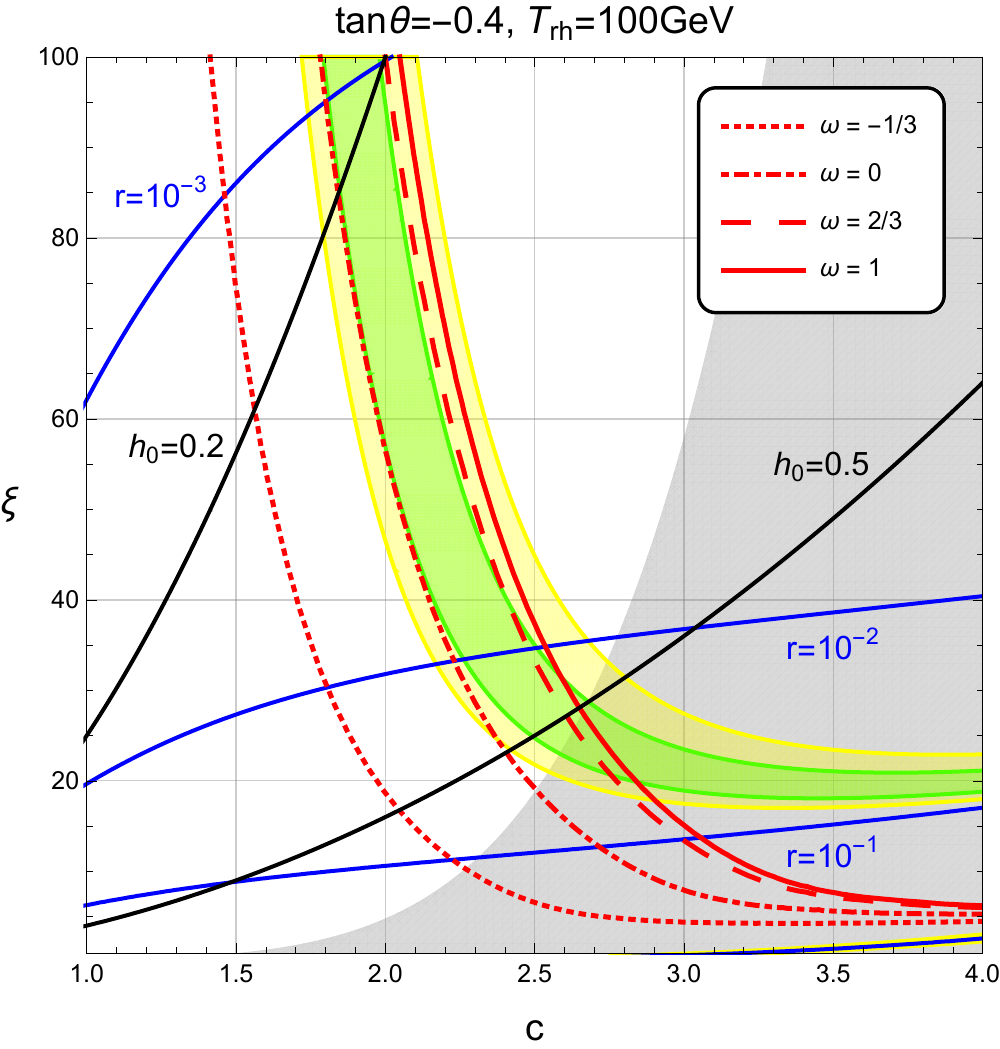} 
    \includegraphics[height=0.30\textwidth]{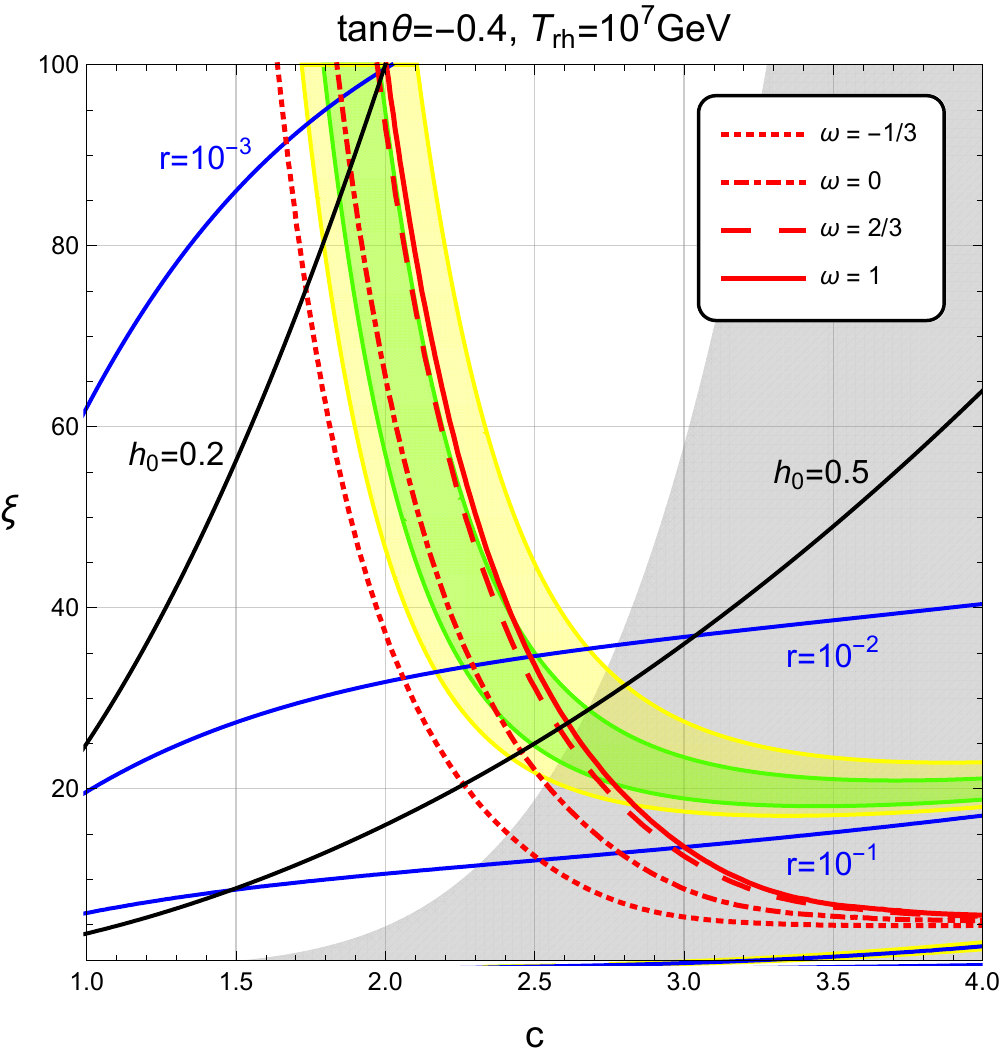}
     \includegraphics[height=0.30\textwidth]{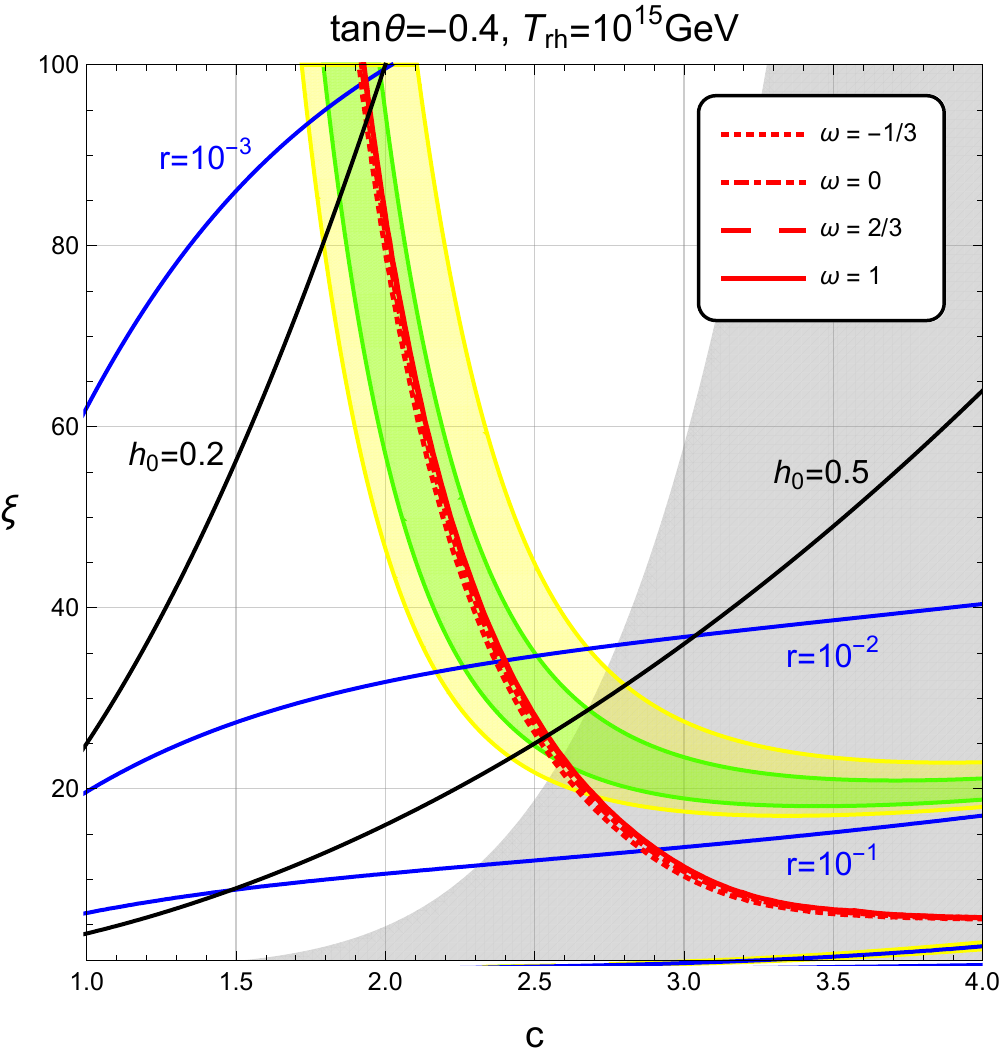}
         \end{center}
  \caption{Parameter space for the inflection point $c=h_0\sqrt{\xi}$ and the non-minimal coupling $\xi$ in the SM or $B-L$ Higgs inflation, for a chosen $\tan\Theta=0, -0.1, -0.4$ from top to bottom panels. Reheating temperature is taken to $T_{\rm rh}=100\,{\rm GeV}, 10^7\,{\rm GeV}, 10^{15}\,{\rm GeV}$, from left to right plots in each panel and the equation of state is chosen to $w=-\frac{1}{3}, 0, \frac{2}{3}, 1$ in red lines in each plot. Others are the same as in Fig.~\ref{Higgs-parameter}.
  }
  \label{reheat-w}
\end{figure}

In Fig.~\ref{reheat-w}, we present the parameter space for the inflection point $c=h_0\sqrt{\xi}$ vs the non-minimal coupling $\xi$ in red lines, for a fixed $\tan\Theta=0,-0.1,-0.4$, from top to bottom panels. In each plot, we take the equation of state $w$ during inflation between $-\frac{1}{3}, 0, \frac{2}{3}, 1$, in dotted, dot-dashed, dashed and solid red lines, and the Planck $1\sigma$ and $2\sigma$ bounds on the spectral index are shown in green and yellow regions, respectively. We vary the reheating temperature to be $T_{\rm rh}=100\,{\rm GeV}, 10^7\,{\rm GeV}, 10^{15}\,{\rm GeV}$ from left to right plots in each panel. The lower the reheating temperatures, the smaller the required number of efoldings at horizon exit becomes for $w<\frac{1}{3}$, opening the wider parameter space for $\xi$ and $c$ in the region with smaller values of $c$. 
Several values of inflection point $h_0$ are shown in each plot, showing that the inflaton field values in Jordan frame are sub-Planckian in most of parameter space. 
We also show the tensor-to-scalar ratio with $r=10^{-1}, 10^{-1.5}$, $r=10^{-1}, 10^{-2}$, and $r=10^{-1}, 10^{-2}, 10^{-3}$, in blue lines, in the upper, middle and lower panels, respectively.

For a small $\tan\Theta$ as in the upper panel of Fig.~\ref{reheat-w}, the tensor-to-scalar ratio tends to be large, saturating the Planck bound, and the required non-minimal coupling is insensitive to the change of the equation of state, in the region favored by the measured spectral index. 
On the other hand, for a sizable $\tan\Theta$ as in the lower panel, the tensor-to-scalar ratio tends to be small, but the predictions for the spectral index and tensor-to-scalar ratio are sensitive to the equation of state in the case of low reheating temperature. In this case, the required non-minimal coupling varies in a wide range of values.

\begin{figure}
  \begin{center}
\includegraphics[height=0.35\textwidth]{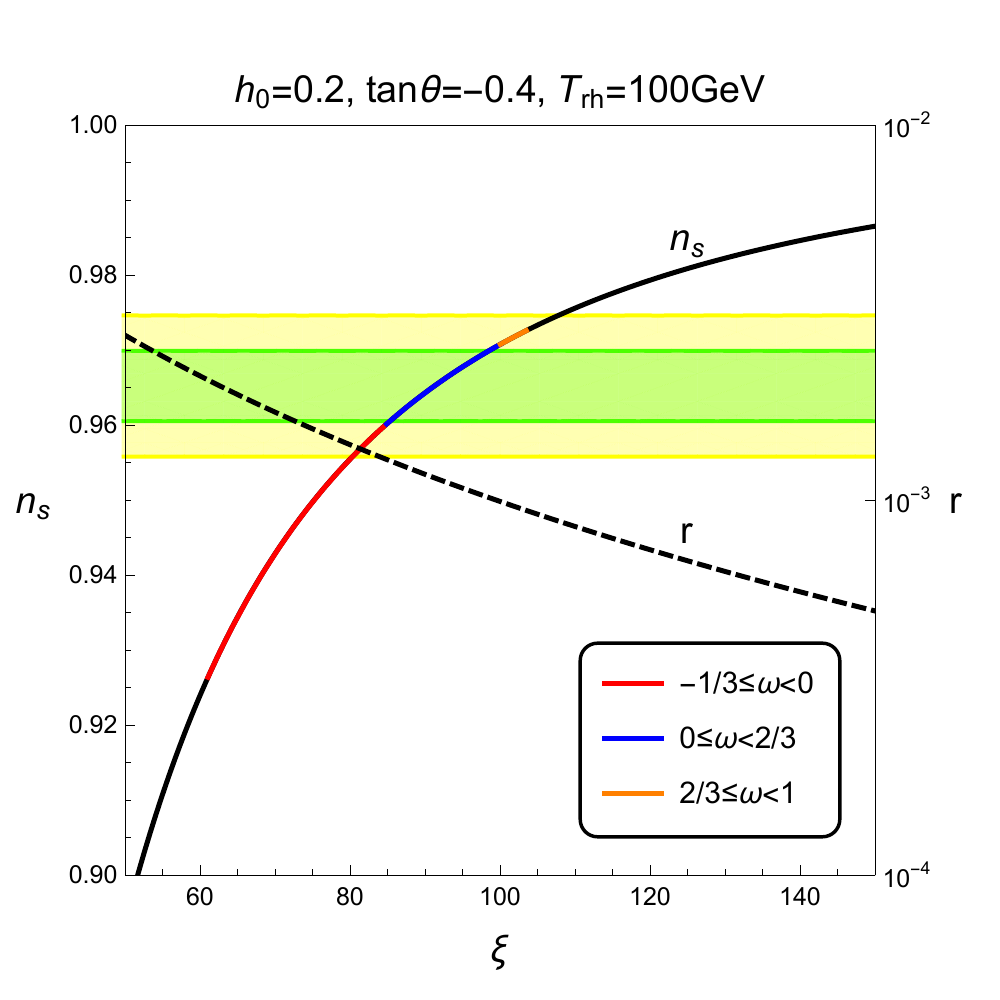}
 \includegraphics[height=0.35\textwidth]{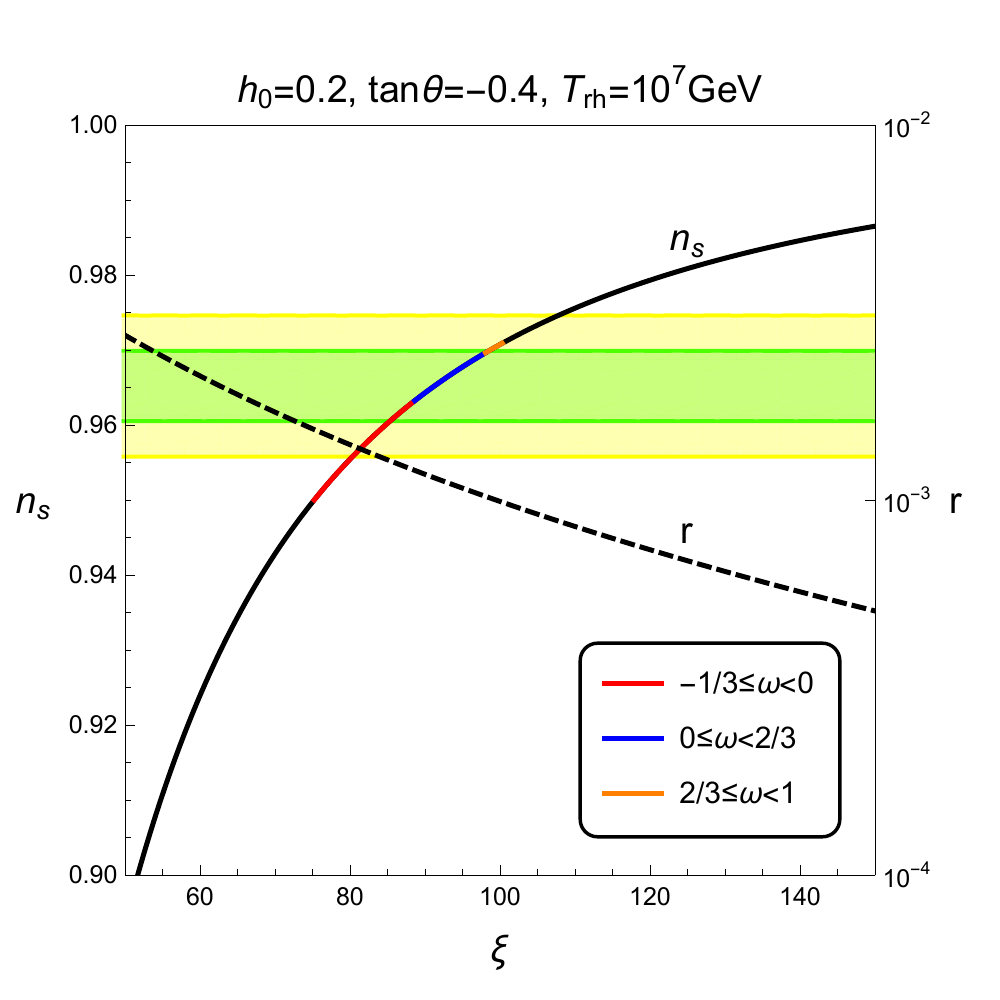} 
  \\
   \includegraphics[height=0.35\textwidth]{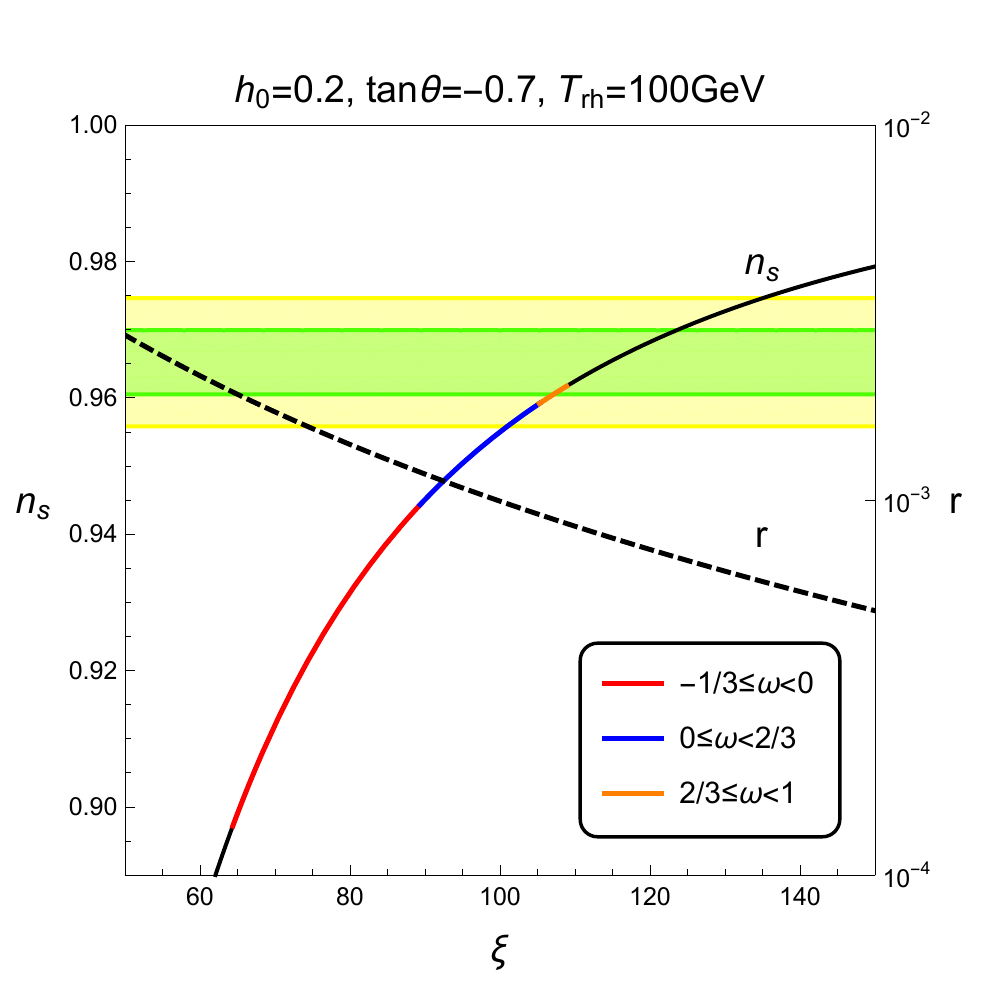}
      \includegraphics[height=0.35\textwidth]{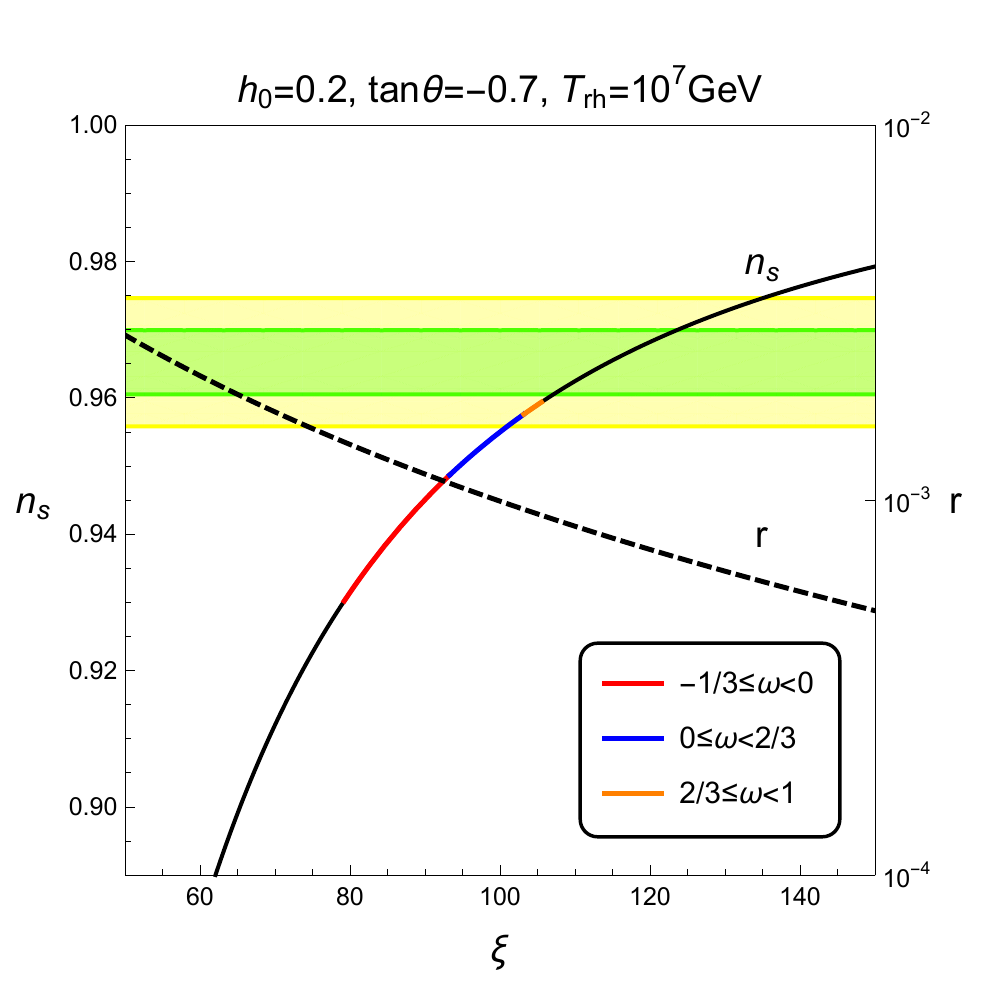}   \\
     \includegraphics[height=0.35\textwidth]{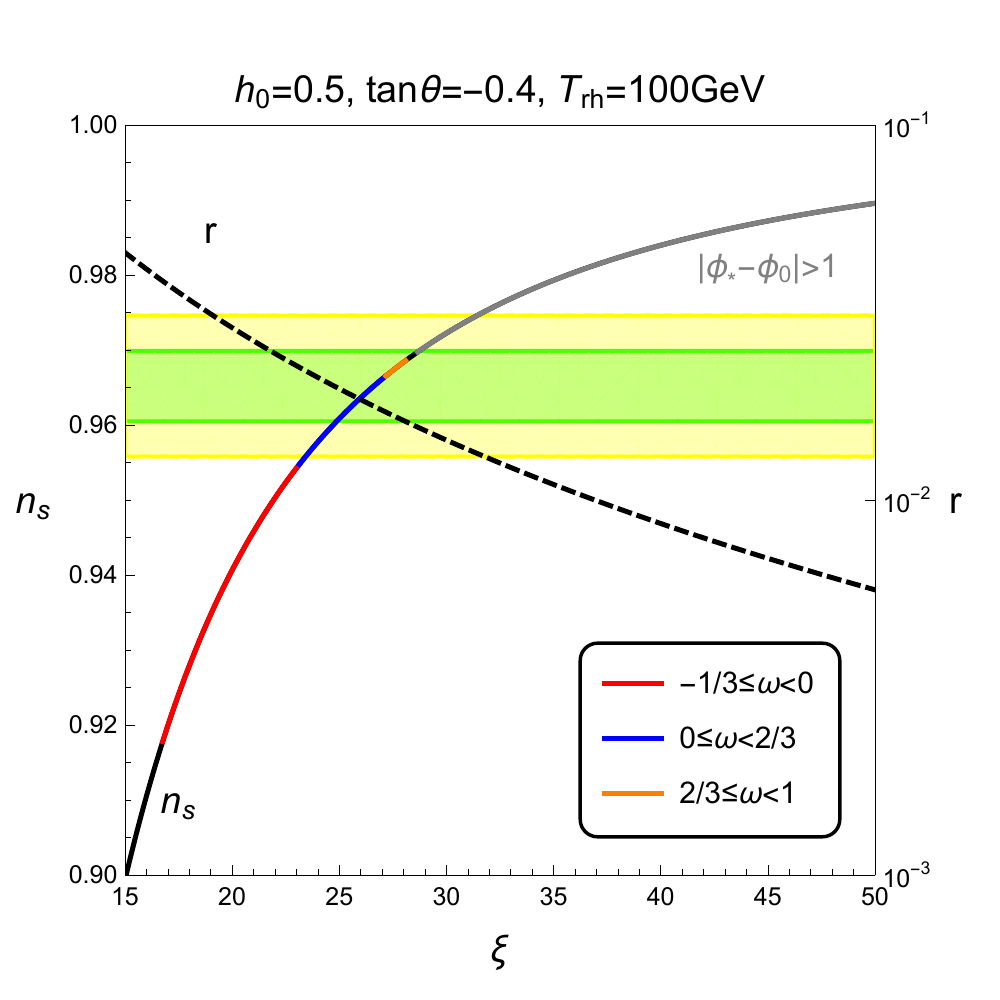}
   \includegraphics[height=0.35\textwidth]{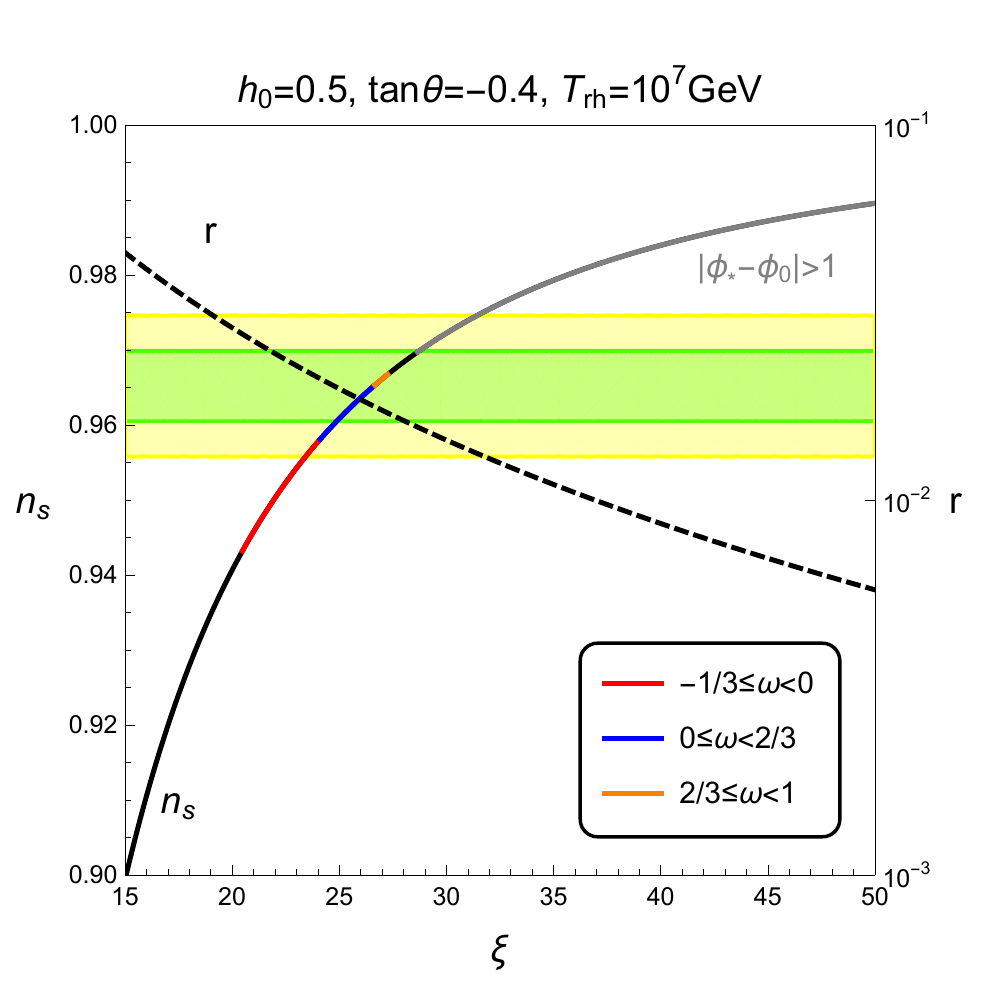}  
           \end{center}
  \caption{The spectral index as a function of the non-minimal coupling $\xi$ in the SM or $B-L$ Higgs inflation, given a set of the inflection point $h_0$ and $\tan\Theta$, $(0.2, -0.4), (0.2, -0.7), (0.5, -0.4)$, from top to bottom panels.  Reheating temperature is chosen to $T_{\rm rh}=100\,{\rm GeV},  10^7\,{\rm GeV}$, from left to right figures in each panel.  The equation of state during reheating is chosen as $-\frac{1}{3}\leq w<0$, $0\leq w<\frac{2}{3}$, $\frac{2}{3}\leq w<1$, in red, blue and orange lines, in each plot.  Others are the same as in Fig.~\ref{Higgs-ns}. 
}
  \label{Higgs-reheat}
\end{figure}

In Fig.~\ref{Higgs-reheat}, we also show that the spectral index and the tensor-to-scalar ratio as a function of the non-minimal coupling, for a set of inflection point $h_0$ and $\tan\Theta$ to $(0.2, -0.4), (0.2, -0.7), (0.5, -0.4)$, from top to bottom panels. 
The reheating temperature is chosen to $T_{\rm rh}=100\,{\rm GeV}, 10^7 \,{\rm GeV}$, from left to right figures, in each panel. The spectral index varies along the black line, depending on the equation of state, $-\frac{1}{3}\leq w<0$, $0\leq w<\frac{2}{3}$, $\frac{2}{3}\leq w<1$, in red, blue and orange lines in each plot. The region with inflation field value taking $|\phi_*-\phi_0|>1$ at horizon exit is shown in gray line in the lower panel.
As the equation of state during reheating varies from $w=-\frac{1}{3}$ to $w=1$, the spectral index increases and falls into the region favored by Planck while the tensor-to-scalar ratio decreases.
On the other hand, as the reheating temperature decreases, the parameter space for small $w$ widens, but it tends to be disfavored by the measured spectral index in Fig.~\ref{Higgs-reheat}. However, the effect of the decreasing equation of state at low reheating temperature can be compensated by taking a smaller inflection point, as shown in Fig.~\ref{reheat-w}, such that the consistent spectral index is obtained.

\section{Conclusions}

We have considered the inflection point inflation as one of the slow-roll inflation models that might be insensitive to trans-Planckian problems usually encountered in large field inflation models. 
We showed that the inflection point inflation predicting a small tensor-to-scalar ratio is sensitive to the reheating temperature and the equation of state during inflation, whereas the case with a large tensor-to-scalar ratio tends to be robust and insensitive to the reheating dynamics.  
From the examples of the SM Higgs inflation and the $B-L$ Higgs inflation where the reheating temperature is determined,  we presented the inflationary predictions for the spectral index and the tensor-to-scalar ratio, depending on the reheating dynamics. 
In particular, in the $B-L$ Higgs inflation, when the light inflaton has a small Higgs-portal coupling, the reheating temperature is small, so the reheating dynamics could affect the inflationary predictions much if the equation of state deviates from radiation.

\section*{Acknowledgments}

The work of HML is supported in part by Basic Science Research Program through the National Research Foundation of Korea (NRF) funded by the Ministry of Education, Science and Technology (2013R1A1A2007919). The work of SMC is supported by the Chung-Ang University Graduate Research Scholarship in 2016.

\def\theequation{A.\arabic{equation}}

\setcounter{equation}{0}

\vskip0.8cm
\noindent
{\Large \bf Appendix: Two-loop renormalization group equations in the $U(1)_{B-L}$ extension} 
\vskip0.4cm
\noindent

In the $U(1)_{B-L}$ extension of the SM, the quartic couplings in the scalar potential are given by
\be
V_{\rm quartic} = \lambda_H |H|^4 + \lambda_\phi |\phi|^4 + \lambda_{H\phi} |\phi|^2 |H|^2,
\ee
and the Yukawa couplings for the RH neutrinos are ${\cal L}_Y=-\frac{1}{2} y_N \phi \overline {N^c} N+{\rm h.c.}$. 
Then, when three RH Yukawa couplings are equal and the Higgs-portal coupling $\lambda_{H\phi}$ and the gauge kinetic mixing between the SM hypercharge and $U(1)_{B-L}$ gauge bosons are small, the RG equations for the $B-L$ Higgs quartic coupling, the RH neutrino Yukawa couplings and the $B-L$ gauge coupling are given at two-loop order \cite{chun,okada} by
\bea
\frac{d\lambda_\phi}{d\ln\mu} &=&\beta^{(1)}_{\lambda_\phi} + \beta^{(2)}_{\lambda_\phi},\\
\frac{d y_N}{d\ln\mu} &=&\beta^{(1)}_{y_N} + \beta^{(2)}_{y_N},\\
\frac{d g_{B-L}}{d\ln\mu} &=& \beta^{(1)}_{g_{B-L}} + \beta^{(2)}_{g_{B-L}}
\eea 
where one-loop and two-loop beta functions for the $B-L$ Higgs quartic coupling are, respectively, 
\bea
\beta^{(1)}_{\lambda_\phi} &\equiv &\kappa (20\lambda^2_\phi - 3y^4_N+96 g^4_{B-L}+6\lambda_\phi y^2_N-48\lambda_\phi g^2_{B-L}),   \label{lamphi1} \\ 
\beta^{(2)}_{\lambda_\phi} &\equiv&\frac{2\kappa^2}{5} \Big(75 \lambda_\phi g^2_{B-L}y^2_N+90 g^2_{B-L} y^4_N+480 g^4_{B-L} y^2_N+5280 \lambda_\phi g^4_{B-L}+1120 \lambda^2_\phi g^2_{B-L} \nonumber \\
&&-17920 g^6_{B-L}+\frac{45}{2}\lambda_\phi y^4_N -150\lambda^2_\phi y^2_N +270 y^6_N -600\lambda^3_\phi \Big)
\label{lamphi2}
\eea
with $\kappa\equiv 1/(4\pi)^2$, and one-loop and two-loop beta functions for the RH neutrino Yukawa couplings are, respectively,
\bea
\beta^{(1)}_{y_N} &=& \kappa\Big( \frac{5}{2} y^3_N - 6y_N g^2_{B-L} \Big),  \\
\beta^{(2)}_{y_N}  &=& \frac{\kappa^2}{20} \Big(1030 g^2_{B-L}y^3_N +(2540 g^4_{B-L}+45 y^4_N-80 \lambda^2_\phi)y_N \nonumber \\
&&-160 \lambda_\phi y^3_N-10y^5_N \Big),
\eea
and lastly, one-loop and two-loop beta functions for the $B-L$ gauge coupling are, respectively, 
\bea
\beta^{(1)}_{g_{B-L}} &=&  12\kappa g^3_{B-L}, \\
\beta^{(2)}_{g_{B-L}} &=& \frac{\kappa^2}{100}\Big(\frac{80000}{9} g^5_{B-L}+\frac{1840}{3} g^2_1 g^3_{B-L}+1200 g^2_2 g^3_{B-L} \nonumber \\  
&&+\frac{3200}{3} g^2_3 g^3_{B-L}-\frac{400}{3} y^2_t g^3_{B-L}-300y^2_N g^3_{B-L} \Big)
\eea
with $g_{1,2,3}$ being the SM gauge couplings and $y_t$ is the top Yukawa coupling.

\end{document}